\documentclass{jfm}
\usepackage{upmath}
\usepackage{amsmath}
\usepackage{amsfonts}
\usepackage{amssymb}
\usepackage{pgf,pgfarrows}
\usepackage{wasysym}
\usepackage{graphicx}
\usepackage{subfig}
\usepackage{natbib}
\usepackage{upgreek}
\usepackage{mathrsfs}
\usepackage{float}
\usepackage{caption}
\usepackage{soul}

\ifCUPmtlplainloaded \else
  \checkfont{eurm10}
  \iffontfound
    \IfFileExists{upmath.sty}
      {\typeout{^^JFound AMS Euler Roman fonts on the system,
                   using the 'upmath' package.^^J}
}
      {\typeout{^^JFound AMS Euler Roman fonts on the system, but you
                   dont seem to have the}
       \typeout{'upmath' package installed. JFM.cls can take advantage
                 of these fonts,^^Jif you use 'upmath' package.^^J}
       
      }
  \else
    
  \fi
\fi
\ifCUPmtlplainloaded \else
  \checkfont{msam10}
  \iffontfound
    \IfFileExists{amssymb.sty}
      {\typeout{^^JFound AMS Symbol fonts on the system, using the
                'amssymb' package.^^J}
\let\le=\leqslant  \let\leq=\leqslant
         \let\geq=\geqslant
      }{}
  \fi
\fi
\ifCUPmtlplainloaded \else
  \IfFileExists{amsbsy.sty}
    {\typeout{^^JFound the 'amsbsy' package on the system, using it.^^J}
}
    {\providecommand\boldsymbol[1]{\mbox{\boldmath $##1$}}}
\fi

\def\ii{{\rm i}}

\affiliation{
$^1$Institute of Applied Mathematics, University
of British Columbia, Vancouver,
BC, V6T 1Z2, Canada.\\[\affilskip]
$^2$Department of Civil Engineering, University
of British Columbia, Vancouver,
BC, V6T 1Z4, Canada.}
\pubyear{2012}
\volume{xxx}
\pagerange{xxx--yyy}
\title[Wave Interaction Theory]{A wave interaction approach to studying non-modal homogeneous and stratified shear instabilities}
\author[A.~Guha and G.A.~Lawrence]
{A\ls N\ls I\ls R\ls B\ls A\ls N\ns G\ls U\ls H\ls A$^{1,2}$% 
\thanks{Present Address: Atmospheric and Oceanic Sciences, University
of California Los Angeles, Los Angeles, CA 90095-1565.}
\and
\ns G\ls R\ls E\ls G\ls O\ls R\ls Y\ns A.\ns L\ls A\ls W\ls R\ls E\ls N\ls C\ls E$^{2}$}
\date{?? and in revised form ??}

\begin{document}

\maketitle

\begin{abstract}
\cite{holm1962} postulated that resonant interaction between two or more progressive, linear interfacial waves produces exponentially growing instabilities in idealized (broken-line profiles), homogeneous or density stratified, inviscid shear layers. Here we have generalized Holmboe's mechanistic picture of linear shear instabilities by (i) not initially specifying the wave type, and (ii) by providing the option for non-normal growth. We have demonstrated the mechanism behind linear shear instabilities by proposing a \emph{purely} kinematic model consisting of two linear, Doppler-shifted, progressive interfacial waves moving in opposite directions. Moreover, we have found a   \emph{necessary} and \emph{sufficient} (N\&S) \emph{condition}  for the existence of exponentially growing  instabilities in idealized shear flows. The two interfacial waves, starting from \emph{arbitrary} initial conditions, eventually phase-lock  and resonate (grow exponentially), provided the N\&S condition is satisfied. The theoretical underpinning of our wave interaction model is analogous to that of synchronization between two coupled harmonic oscillators. We have re-framed our model into a non-linear autonomous dynamical system, the steady state configuration of  which corresponds to the resonant configuration of the wave-interaction model.  When interpreted in terms of the canonical normal-mode theory, the  steady state/resonant configuration corresponds to the growing normal-mode of the discrete spectrum. The instability mechanism occurring prior to reaching steady state is  non-modal,  favouring rapid transient growth. Depending on the wavenumber and initial phase-shift,  non-modal gain can exceed the corresponding modal gain by many orders of magnitude. Instability is also observed in the parameter space which is deemed stable by the normal-mode theory.  Using our model we have derived the   discrete spectrum non-modal stability equations for three classical examples of shear instabilities  - Rayleigh/Kelvin-Helmholtz, Holmboe and Taylor-Caulfield. We have shown that the N\&S condition provides a range of unstable wavenumbers for each instability type, and this range matches the predictions of the normal-mode theory. 

\end{abstract}

\section{Introduction}

Statically stable density stratified shear layers are ubiquitous in  the atmosphere and oceans. Such shear layers can become hydrodynamically unstable, leading to turbulence and mixing in geophysical flows.
Turbulence and mixing strongly influence the atmospheric and oceanic circulation - processes known to play key roles in shaping our weather and  climate. Hydrodynamic instability is characterized by the
growth of wavelike perturbations in a laminar base flow. Such perturbations can grow at an exponential rate, transforming the base flow from a laminar to a turbulent state. In the present study, we will   theoretically investigate the underlying mechanism(s) leading to modal (exponential), as well as non-modal, growth of small wavelike perturbations in idealized homogeneous and stratified shear flows.

The classical method used to determine flow stability is the normal-mode approach of linear stability analysis \cite[]{draz1966,draz1982}. Under the normal-mode assumption, a  waveform can grow or decay
but cannot deform. Normal-mode perturbations are added to the laminar background flow profile, followed by linearizing  the governing Navier Stokes equations. 
For inviscid, density stratified shear flows, 
the normal-mode formalism leads to the celebrated Taylor-Goldstein equation, derived independently by \cite{tayl1931} and \cite{gold1931}. The Taylor-Goldstein equation is  an eigenvalue problem which
calculates the wave properties like growth-rate, phase-speed, and eigenfunction associated with each normal-mode. For stability analysis, the range of unstable wavenumbers  and the wavenumber
corresponding to the fastest growing normal-mode are of prime interest. %In many occasions, the fastest growing mode dominates over all other modes, therefore it governs the behaviour of the transitional flow.
 In many practical scenarios, the Taylor-Goldstein equation
is found to accurately capture the onset of instability \cite[]{thor1973}, and it also provides a first order description of the developing flow structures \cite[]{law1991,ted2009}. 

The normal-mode approach, however, has  shortcomings. Firstly, it provides limited insight into the physical mechanism(s) responsible for hydrodynamic instability. 
The answer to why an infinitesimal perturbation vigorously grows in an otherwise stable background flow is provided in the form of mathematical theorems - Rayleigh-Fj$\o$rtoft's theorem  for homogeneous flows,
and Miles-Howard criterion for stratified flows \cite[]{draz1982}.  It is often difficult to form an intuitive understanding of shear instabilities from these theorems. Since linear instability is the first step towards understanding the more complicated and highly elusive non-linear processes like chaos and turbulence, it is desirable to explore alternative
routes in order to provide additional insight. In this context Sir G. I. Taylor writes: ``It is a simple matter to work out with the equations which must be satisfied by waves in such a fluid, but the interpretation of the solutions of these equations is a matter of considerable difficulty''  \cite[]{tayl1931}. 
 A second drawback of the normal-mode approach is the normal-mode assumption itself.  The extensive work by \cite{farrell1984modal}, \cite{tref1993}, \cite{schmid2001stability}
and others have shown that shear allows rapid non-modal transient growth due to non-orthogonal interaction between the modes. \cite{farrell1996} 
 developed the ``Generalized stability theory''  for obtaining the optimal non-modal growth from the singular value decomposition of the propagator matrix of a linear dynamical system.

  Lord Rayleigh  was probably the first to inquire about the mechanism behind shear instabilities, and conjectured the possible role of wave interactions \cite[]{rayl1880}. Lord Rayleigh's hypothesis was later corroborated by Sir G. I. Taylor while he was theoretically studying three-layered flows in constant shear. He explained the mechanism to be as follows:  ``Thus the instability  might  be
regarded as being due to a free wave at the lower surface  forcing a  free wave at the upper surface of separation when their velocities in space coincide'' \cite[]{tayl1931}. Early explorations of the wave interaction concept have been reviewed in \cite{draz1966}.  The first mathematically concrete mechanistic description of stratified shear instabilities was provided by \cite{holm1962}. 
Using idealized velocity and density profiles, Holmboe postulated that the resonant interaction between stable propagating waves, each existing at 
a discontinuity in the background flow profile (density profile discontinuity produces interfacial gravity waves and vorticity profile discontinuity produces vorticity waves),
yields exponentially growing instabilities. He was able to show that Rayleigh/Kelvin-Helmholtz (hereafter, ``KH'') instability  \cite[]{rayl1880} is the result of the interaction between two vorticity waves (also known as Rayleigh waves). Moreover, Holmboe  also  found a new 
type of instability, now known as the ``Holmboe instability'',  produced by the interaction between vorticity and interfacial gravity waves. Bretherton, a contemporary of Holmboe, proposed a similar theory to explain mid-latitude cyclogenesis \cite[]{bret1966}. 
He hypothesized that cyclones form due to a baroclinic instability  
caused by the interaction between two Rossby edge waves (vorticity waves in a rotating frame of reference), one existing at the earth's surface and the other located at the atmospheric tropopause. 

The theories proposed by Holmboe and Bretherton have been refined and re-interpreted over the years, see \cite{cair1979,hosk1985,caul1994,bain1994,heif1999,carp2012}.  
As reviewed in \cite{carp2012}, resonant interaction between two edge waves in an idealized homogeneous or stratified shear layer occurs when these waves attain a phase-locked state, i.e.\ they are at  rest relative to each other. Maintaining this phase-locked configuration, the waves grow equally at an exponential rate. 
 There is also an alternative description of shear instabilities through wave interactions which was put forward by \cite{cair1979}. He
introduced  the concept of ``negative energy waves'', which
are stable modes, and their introduction into the flow causes a decrease in the total energy. Whether a given wave mode has positive or negative energy
depends on the frame of reference used.
Instability  results when negative energy mode resonates with a positive
energy mode, and this occurs when the waves have the same phase-speed and wavelength. This can
be identified by the crossing of dispersion curves for the positive and negative energy modes in a frequency-wavenumber diagram. Yet another 
 mechanistic picture of shear instabilities was proposed by Lindzen 
and co-authors (summarized in \cite{lind1988}). This theory, known as the ``Over-reflection theory'', proposes that under the right flow configuration, over-reflection
of waves can continuously energize an advective {}``Orr process'' \cite[]{orr1907} which is finally responsible for the perturbation growth. 

The present paper focusses on studying shear instabilities in terms of wave interactions. From the recent review by  \cite{carp2012} it can be inferred that the wave interaction theory is  in its early phases of  development. In fact, there is no strong theoretical justification behind the argument that two progressive waves lock in phase and resonate, thereby producing exponentially growing instabilities in a shear layer. Many questions in this context remains unanswered, e.g. (a) Is there a condition which determines whether two waves will lock in phase? (b) Starting from an initial condition, how long  does it take for the waves to get phase-locked? (c) If exponentially growing instabilities occur after phase-locking, then what kind of instabilities (if any) occur prior to phase-locking? A point worth mentioning here is that the phenomenon of phase-locking occurs in diverse problems ranging from biology to electronics \cite[]{PIK01}. In fact ``synchronization'' is an area of study (in dynamical systems theory) specifically dedicated to this purpose. The history of synchronization goes back to the $17$th century when the famous Dutch scientist Christiaan Huygens reported his observation of synchronization of two pendulum clocks.  
We suspect that there may be an analogy between the fundamental aspects of synchronization theory and that of wave interaction based interpretation of shear instabilities. If  exploited successfully, this analogy could be beneficial in answering some of the key questions concerning the origin and evolution of shear instabilities.  

In recent years, Heifetz and co-authors \cite[]{heif1999,heifetz2004counter,heif2005} have extensively studied the interaction between Rossby edge waves. They performed a detailed analysis of non-modal instability and transient growth mechanisms in idealized barotropic shear layers. %In fact Heifetz and co-authors have answered some of the questions we raised in the previous paragraph, but their analysis is only applicable to the case of (idealized) barotropic shear instability. 
Modal and non-modal growth occurring due to the interactions between surface gravity waves and a pair of vorticity waves (existing at the interfaces of a submerged piecewise linear shear layer) was recently studied by \cite{bakas2009}.

The goal of our paper is to frame a  theoretical model of shear instabilities, which could be applied to different idealized (broken-line) shear layer profiles (e.g. the classical profiles studied by Rayleigh, Taylor or Holmboe). The model should be sufficiently  generalized in the sense that the constituent wave types need not be specified \emph{a priori}.  The model should be able to capture the transient dynamics, hence should not  be limited to normal-mode waveforms.
%By \emph{not} limiting our analysis to normal-mode waveforms, and furthermore, by \emph{not} 
%assuming any particular type of waveform (e.g.\  gravity wave or vorticity wave), we have formulated a generalized wave interaction model of shear instabilities. 
%Our equations are derived from the linearized kinematic and dynamic (for stratified flows) 
%conditions. Unlike \cite{heifetz2004counter} and \cite{heif2005}, our key variables are the vertical displacement and  the vertical velocity of the wave. The choice of variables, along with the derivation methodology, provide a deeper
%and probably more intuitive understanding of the wave interaction process. 

The outline of the paper is as follows. In \S \ref{sec:linlin}
we  provide a theoretical background of linear progressive waves, and focus on two  types of waves - vorticity waves and internal gravity waves. The wave theory in this section 
is more generalized than that usually reported in the literature. In \S \ref{sec:waveint} we  investigate the mechanism of interaction between two progressive waves. We undertake a dynamical systems approach to
better understand the wave interaction problem, especially the resonant condition. Since the wave interaction formulation is not restricted to normal-mode type instabilities, we investigate non-modal/transient growth processes in \S \ref{sec:matstuff}. Finally in \S \ref{sec:inst}  we use wave interactions to analyze three well known types of shear instabilities   - KH 
instability (resulting from the interaction between two vorticity waves), Taylor-Caulfield instability (resulting from the interaction
between two interfacial internal gravity waves), and Holmboe instability (resulting from the interaction between a vorticity wave and an interfacial internal gravity wave). 
\section{Linear wave(s) at an interface}
\label{sec:linlin}
Consider a fluid interface existing at $z=z_{i}$ in an unbounded, inviscid, incompressible, two dimensional ($x$-$z$) flow. Moreover, let the interface be perturbed by an infinitesimal displacement $\eta_{i}$ in the $z$ direction, given as follows:
%In the present study we consider multi-layered flows with constant  density and vorticity in each layer.  This configuration makes  the equations of motion for perturbations within a layer to be the same as that in an irrotational background flow.
%The interface between two adjacent layers signifies a discontinuity in vorticity or density. The former is a vorticity interface,
%while the latter is a density interface.
%Let such an interface existing at a location $z=z_{i}$ be perturbed by an infinitesimal vertical displacement  $\eta_{i}$, given as follows:
 \begin{equation}
\eta_{i} = \Re \{ A_{\eta_{i}}(t) e^{\ii [\alpha x+\phi_{\eta_{i}}(t)]} \}.
\label{eq:eta}
\end{equation}
This displacement manifests itself in the form of stable progressive wave(s), the amplitude and phase of which are $A_{\eta_{i}}$ and $\phi_{\eta_{i}}$, respectively.
For example, when this interface is a vorticity interface, it produces a vorticity wave. Likewise, two oppositely traveling gravity waves
are produced in case of a density interface. We have assumed the interfacial displacement (or the wave) to be monochromatic, having a wavenumber $\alpha$. Moreover, the interface satisfies the \emph{kinematic condition} -  a particle initially on the interface will remain there forever. The linearized kinematic condition is given by
\begin{equation}
\frac{\partial\eta_{i}}{\partial t}+U_{i}\frac{\partial\eta_{i}}{\partial x}=w_{i},
\label{eq:kincon}
\end{equation}
where $U_{i} \equiv U(z_{i})$ is the background velocity in the $x$ direction and $w_{i}$ is the $z$-velocity at the interface. We prescribe the latter to be as follows:
\begin{equation}
w_{i}=\Re\{A_{w_{i}}(t) e^{\ii[\alpha x+  \phi_{w_{i}}(t)]}\}.
\label{eq:wi}
\end{equation}
Here $A_{w_{i}}$ is the amplitude and $\phi_{w_{i}}$ is the phase of $w_{i}$. 
The interfacial displacement creates  vorticity perturbation \emph{only} at the interface, the perturbed velocity field is irrotational everywhere else in the domain. Thus 
\begin{equation}
\frac{\partial^{2}\psi}{\partial x^{2}}+\frac{\partial^{2}\psi}{\partial z^{2}}=0\,\,\,\mathrm{when} \,\,\, z  \neq z_{i}. 
\label{eq:lapup}
\end{equation}
 In the above equation $\psi$ is the perturbation streamfunction. Assuming $\psi(x,z,t)=\Re\{\varphi(z)\exp[\alpha x+\phi_{\psi}(t)]\}$, and substituting it in  (\ref{eq:lapup}) we get 
\begin{equation}
\frac{\partial^{2}\varphi}{\partial z^{2}}-\alpha^{2}\varphi=0.
\end{equation}
The above equation yields $\varphi=e^{-\alpha\left|z-z_{i} \right|}\varphi_{i}$ (where $\varphi_{i} \equiv \varphi(z_{i})$). The vertical velocity $w=-\partial \psi / \partial x$ is then given by 
\begin{equation}
w=e^{-\alpha\left|z-z_{i} \right|}w_{i}.
\label{eq:w}
\end{equation}
Thus, like the streamfunction, the vertical velocity  decays exponentially away from the interface, and vanishes at infinity. 

Substituting (\ref{eq:eta}) and (\ref{eq:wi}) in  (\ref{eq:kincon}), we obtain
\begin{equation}
\dot{A}_{\eta_{i}}\cos\left(\alpha x+\phi_{\eta_{i}}\right)-A_{\eta_{i}}\left(\alpha U_{i}+\dot{\phi}_{\eta_{i}}\right)\sin\left(\alpha x+\phi_{\eta_{i}}\right)=
A_{w_{i}}\cos\left(\alpha x+\phi_{w_{i}}\right).
 \label{eq:singwave}
\end{equation} 
  Here $\phi_{\eta_{i}},\,\phi_{w_{i}}\in\left[-\pi,\,\pi\right]$.
  The frequency and the growth-rate of a wave are respectively defined as $\Omega_{i}\equiv -\dot{\phi}_{\eta_{i}}$ and  $\gamma_{i} \equiv \dot{A}_{\eta_{i}}/A_{\eta_{i}}$ (overdot denotes $d/dt$). 
  Using these definitions in  (\ref{eq:singwave}), we get % \footnote{In order to obtain  (\ref{eq:1freq})-(\ref{eq:13}) from  (\ref{eq:singwave}), we write the R.H.S. of   (\ref{eq:singwave}) as follows: $A_{w_{i}}\cos\left(\alpha x+\phi_{w_{i}}\right)=A_{w_{i}}\cos\left(\alpha x+\phi_{\eta_{i}}+\triangle\phi_{ii}\right)$. The cosine function is expanded using 
%a standard trigonometric identity. Finally we collect the coefficients of $\sin(\alpha x)$ and $\cos(\alpha x)$.}
\begin{eqnarray}
 &  & \Omega_{i}=\alpha U_{i}-\upomega_{i}\sin\left(\triangle\phi_{ii}\right) \label{eq:1freq} \\
 &  & \gamma_{i}=\upomega_{i}\cos\left(\triangle\phi_{ii}\right), \label{eq:13} 
% &  & c_{i}=U_{i}-\frac{\upomega_{i}}{\alpha}\sin\left(\triangle\phi_{ii}\right) \label{eq:14}
\end{eqnarray}
where  $\triangle\phi_{ii} \equiv \phi_{w_{i}}-\phi_{\eta_{i}}$.  %, and $\upomega_{i} \equiv A_{w_{i}}/A_{\eta_{i}}$ is the magnitude of the \emph{intrinsic} frequency of the wave. 
Eq.\ (\ref{eq:1freq}) shows that the frequency of a wave consists of two components - (i) the Doppler
shift $\alpha U_{i}$, and (ii) the \emph{intrinsic frequency}  $\Omega_{i}^{intr}\equiv-\upomega_{i}\sin\left(\triangle\phi_{ii}\right)$, where $\upomega_{i}\equiv A_{w_{i}}/A_{\eta_{i}}$. The phase-speed $c_{i} \equiv \Omega_{i}/\alpha$  of the wave is found to be
\begin{equation}
c_{i}=U_{i}+c_{i}^{intr},
\label{eq:14}
\end{equation}
where $c_{i}^{intr}\equiv-\left(\upomega_{i}/\alpha\right)\sin\left(\triangle\phi_{ii}\right)$ denotes the \emph{intrinsic phase-speed}. Noting that a wave in isolation cannot grow or decay on its own,  (\ref{eq:13}) demands that $\left| \triangle \phi_{ii} \right|=\pi/2$. 
Therefore for a stable wave,
the vertical velocity field at the interface has to be in quadrature with the interfacial deformation. Another point worth mentioning is that an isolated wave cannot accelerate or decelerate on its own, hence $c_{i}^{intr}$ should be constant. This in turn means $\upomega_{i}$ is a constant quantity.

 Applying  the quadrature condition to  (\ref{eq:1freq}) and  (\ref{eq:14}),  the magnitudes of the intrinsic frequency and  the intrinsic phase-speed  become $\left|\Omega_{i}^{intr}\right|=\upomega_{i}$ and $\left|c_{i}^{intr}\right|=\upomega_{i}/\alpha$, respectively.  The \emph{intrinsic direction of motion} of the wave, however, is determined by $\triangle \phi_{ii}$. For waves moving to the left relative to the interfacial velocity ($U_{i}$), $\triangle \phi_{ii}=\pi/2$. Similarly for right moving waves, $\triangle \phi_{ii}=-\pi/2$.  When such a stable progressive wave is acted upon by external influence(s) (e.g.\ when another wave interacts with the given wave, as detailed in \S \ref{sec:waveint}), the quadrature condition is no longer satisfied, i.e.\  $\left| \triangle \phi_{ii} \right| \ne \pi/2$. Therefore, the wave may  grow ($\gamma_{i}>0$) or decay ($\gamma_{i}<0$), and its  intrinsic frequency  $\Omega_{i}^{intr}$ and phase-speed $c_{i}^{intr}$ may change.  
 
 In our analyses we will consider  two types of  progressive interfacial  waves - vorticity waves and internal gravity waves. 

\subsection{Vorticity/Rayleigh Waves}
\label{subsec:vort_waves}
Vorticity waves, also known as Rayleigh waves, exist at a vorticity interface (i.e.\ regions involving a sharp change in vorticity). Such interfaces  are a common feature in the atmosphere and oceans. In a rotating frame, the analog of the vorticity wave is the Rossby edge wave which exists at a sharp transition in the potential vorticity. When Rossby 
edge waves  propagate
in a direction opposite to the background flow, they are called ``counter-propagating Rossby waves''  \cite[]{heif1999}. 

In order to evaluate the frequency  $\upomega_{i}$  of vorticity waves, let us consider a velocity profile having the form
 \begin{equation}
U\left(z\right)=\begin{cases}
U_{i} & z\geq z_{i}\\
S z & z\leq z_{i}. \end{cases}\label{eq:V1}\end{equation} 
Here the constant $S=U_{i}/z_{i}$  is the vorticity, or the shear in the region $z\leq z_{i}$.  Eq.\ (\ref{eq:V1}) shows that the vorticity $dU/dz$ is discontinuous at $z=z_{i}$. 
This condition supports a vorticity wave. An interfacial deformation $\eta_{i}$ adds vorticity $S$ to the upper layer and removes it from the lower layer, thereby creating a vorticity imbalance at the interface. This imbalance is countered by the background vorticity gradient, which acts as a restoring force, and thereby leads to wave propagation. 

The horizontal component $u_{i}$ of the perturbation velocity field  set up by the interfacial deformation undergoes a jump at the interface, the value of which can be determined from Stokes' Theorem (see Appendix \ref{app:B}):
\begin{equation}
u_{i}^{+}-u_{i}^{-}=S\eta_{i}.
\label{eq:Stokes}
\end{equation}
By taking an $x$ derivative of  (\ref{eq:Stokes}) and invoking the continuity relation, we get
\begin{equation}
-\frac{\partial w_{i}^{+}}{\partial z}+\frac{\partial w_{i}^{-}}{\partial z}=S\frac{\partial\eta_{i}}{\partial x}.
\label{eq:vort12}
\end{equation}
%If the interfacial disturbance is sinusoidal and is given by $\eta_{i}= A_{\eta_{i}}(t)\cos\left[\alpha x+\phi_{\eta_{i}}(t)\right]$, 
%the vertical velocity field becomes $w_{i}=e^{-\alpha\left|z-z_{i}\right|} A_{w_{i}}(t)\cos\left[\alpha x+\phi_{w_{i}}(t)\right]$. This means that the quantity $w_{i}$ is continuous at $z=z_{i}$. 
By substituting   (\ref{eq:eta}) and (\ref{eq:w})  in  (\ref{eq:vort12}) we obtain 
\begin{equation}
\upomega_{i} =-\frac{S}{2}\frac{\sin\left(\alpha x+\phi_{\eta_{i}}\right)}{\cos\left(\alpha x+\phi_{w_{i}}\right)}=\frac{S}{2\sin\left(\frac{\pi}{2} \mathrm{sgn}(S)\right)},
%\upomega_{i}=-\frac{S}{2}
\label{eq:vortwaverel}
\end{equation}
%The fact that $\upomega_{i}$ is always  positive  demands 
% \begin{equation}
%\triangle\phi_{ii}=\frac{\pi}{2} \mathrm{sgn}(S),
%\end{equation}
where $\mathrm{sgn}(\,)$ is the sign function. From  (\ref{eq:vortwaverel}) $\Omega_{i}^{intr}$ of a vorticity wave is found to be $-S/2$.
 The phase-speed $c_{i}$ can be evaluated by substituting   (\ref{eq:vortwaverel}) into  (\ref{eq:14}):
 \begin{equation}
c_{i}=U_{i} - \frac{S}{2 \alpha}.
\label{eq:vortspeed}
\end{equation}
 If $S>0$, the vorticity wave moves to the left relative to the background flow.  The conventional derivation of the frequency and phase-speed of a vorticity wave can be found in \cite{suth2010}.
 \subsection{Interfacial Internal Gravity Waves}
 \label{subsec:gravitywaves}

Interfacial gravity waves exist at a density interface, i.e.\  a region involving sharp change in density. The most common example is the surface gravity wave existing at the air-sea interface. However, in this paper we will only consider  waves existing at the interface  between two fluid layers having small density difference. Such waves are known as  
 interfacial \emph{internal} gravity waves (hereafter, gravity waves), and often arise in the pycnocline region of density stratified natural water bodies like  lakes, estuaries and oceans.
% Such waves exist at the interface between fluid layers of two different densities, provided the density difference is small.  Natural water bodies like being density stratified
%NatuFor example, gravity waves 
%Such interfaces commonly occur in density stratified 
% type of feature occurs in regions having density stratified flows having a thin density interface (pycnocline). Since most natural water bodies like lakes, estuaries and oceans
%are density stratified, gravity waves are  ubiquitous. 

 In the case of gravity waves,  $\Omega_{i}^{intr}$ can be evaluated by considering the \emph{dynamic condition}. This condition implies that the pressure at the density interface must be continuous.  Let the density of upper and lower fluids be $\rho_{1}$ and $\rho_{2}$ respectively. The background velocity is constant, and is equal to $U_{i}$. Then the linearized dynamic condition at the interface $z=z_{i}$ after some simplification becomes  ((3.13) of \cite{caul1994}):
\begin{equation}
\frac{\partial\psi_{i}}{\partial t}+U_{i}\frac{\partial\psi_{i}}{\partial x}=\frac{g'}{2\alpha}\frac{\partial\eta_{i}}{\partial x}.
\label{eq:dynbound}
\end{equation}
Here $g' \equiv g(\rho_{2}-\rho_{1})/\rho_{0}$ is the reduced gravity and $\rho_{0}$ is the reference density.  Under Boussinesq approximation $\rho_{0} \approx \rho_{1}  \approx \rho_{2} $. By taking an $x$ derivative of  (\ref{eq:dynbound})
and using the streamfunction relation $\left\{ u_{i},w_{i}\right\} =\left\{ -\partial\psi_{i}/\partial z,\,\partial\psi_{i}/\partial x\right\}$, we get
\begin{equation}
\frac{\partial w_{i}}{\partial t}+U_{i}\frac{\partial w_{i}}{\partial x}=\frac{g'}{2\alpha}\frac{\partial^{2} \eta_{i}}{\partial x^{2}}.
\label{eq:gravrel}
\end{equation}
 Substitution of   (\ref{eq:eta}) and (\ref{eq:wi})   in  (\ref{eq:gravrel})  yields 
\begin{eqnarray}
 &  & \dot{A}_{w_{i}}\cos\left(\alpha x+\phi_{w_{i}}\right)-A_{w_{i}}\dot{\phi}_{w_{i}}\sin\left(\alpha x+\phi_{w{}_{i}}\right)\nonumber \\ 
 &  & -\alpha U_{i}A_{w_{i}}\sin\left(\alpha x+\phi_{w{}_{i}}\right)=-\frac{g'\alpha}{2}A_{\eta_{i}}\cos\left(\alpha x+\phi_{\eta_{i}}\right).
 \label{eq:12221}
\end{eqnarray}
 The quantity $\dot{\phi}_{w_{i}}=\dot{\phi}_{\eta_{i}}=-\Omega_{i}=-\alpha c_{i}$. 
On substituting this relation in (\ref{eq:12221})  we obtain 
  \begin{equation}
\upomega_{i} =\frac{g'}{2\left(U_{i}-c_{i}\right)\sin\left(\frac{\pi}{2}\mathrm{sgn}(U_{i}-c_{i})\right)}.
\label{eq:gravwaverel}
\end{equation}

An important aspect of  (\ref{eq:gravwaverel}) is that it has been derived \emph{independent} of the kinematic condition. The presence of single or multiple interfaces \emph{does not} alter the expression in (\ref{eq:gravwaverel}),
implying that this equation provides a generalized description of $\upomega_{i}$. Inclusion of the kinematic condition yields an expression for $\upomega_{i}$ which is simpler, but problem specific.
 For example, when a  single interface is present, inclusion of the kinematic condition in   (\ref{eq:gravwaverel}) produces the well known expression for gravity wave frequency. This can be shown by 
substituting (\ref{eq:1freq}) (Note that this equation has been derived from (\ref{eq:kincon}), which is the kinematic condition for a single interface.)  in 
 (\ref{eq:gravwaverel}) and considering only the positive value:
  \begin{equation}
\upomega_{i} = \sqrt{\frac{g' \alpha}{2}}.
\label{eq:gravwaverel111}
\end{equation} 
The above equation is the dispersion relation for gravity waves. Substitution of (\ref{eq:gravwaverel}) in  (\ref{eq:14}) produces the well known expression for the phase-speed of a gravity wave:
 \begin{equation}
c_{i}=U_{i} \pm \sqrt{\frac{g'}{2 \alpha}}.
\label{eq:gravspeed}
\end{equation}
The above equation shows that each density interface supports two gravity waves, one  moving to the left and the other to the right relative to the background velocity $U_{i}$.
 
 It is not uncommon in the literature to use the expressions (\ref{eq:gravwaverel111}) and (\ref{eq:gravspeed}) \emph{even} when multiple interfaces are present. Such usage certainly leads to erroneous results, especially when the objective is  to study multiple wave interactions. Probably the confusion arises from the traditional derivation of gravity wave frequency and phase-speed  \cite[]{kundu2004,suth2010}.  This derivation strategy obscures the fact that the kinematic condition (at an interface) is influenced by the number of interfaces present in the flow, while the dynamic condition is not.

 \section{Interaction between  two linear interfacial waves}
\label{sec:waveint}
Let us now consider a system with two interfaces, one at $z=z_{1}$ and the other one
at $z=z_{2}$. The linearized kinematic condition at each of these interfaces is given by:
\begin{eqnarray}
& & \frac{\partial\eta_{1}}{\partial t}+U_{1}\frac{\partial\eta_{1}}{\partial x}=w_{1}+e^{-\alpha\left|z_{1}-z_{2}\right|}w_{2}\label{eq:kin1}\\
& & \frac{\partial\eta_{2}}{\partial t}+U_{2}\frac{\partial\eta_{2}}{\partial x}=e^{-\alpha\left|z_{1}-z_{2}\right|}w_{1}+w_{2}\label{eq:kin2}.
\end{eqnarray}
It has been implicitly assumed that both waves have the same  wavenumber $\alpha$. The R.H.S.\ of  (\ref{eq:kin1})-(\ref{eq:kin2}) reveal the subtle effect of wave interaction, 
and can be understood as follows. 
The effect of $w_{1}$  extends away from the interface $z_{1}$,  hence it can be felt by a wave existing at another location, say $z_{2}$. Therefore the  vertical velocity of the wave at $z_{2}$ gets modified - it  becomes the linear
superposition of its own vertical velocity $w_{2}$ and the component of $w_{1}$ existing at $z_{2}$. This phenomenon is also known as ``action-at-a-distance'', see \cite{heif2005}.

On substituting   (\ref{eq:eta}) and (\ref{eq:wi})  in  (\ref{eq:kin1})-(\ref{eq:kin2}), we get
\begin{eqnarray}
\dot{A}_{\eta_{1}}\cos\left(\alpha x+\phi_{\eta_{1}}\right)-A_{\eta_{1}}\left(\alpha U_{1}+\dot{\phi}_{\eta_{1}}\right)\sin\left(\alpha x+\phi_{\eta_{1}}\right)=\nonumber \\
A_{w_{1}}\cos\left(\alpha x+\phi_{w_{1}}\right)+e^{-\alpha\left|z_{1}-z_{2}\right|}A_{w_{2}}\cos\left(\alpha x+\phi_{w_{2}}\right) \\
\dot{A}_{\eta_{2}}\cos\left(\alpha x+\phi_{\eta_{2}}\right)-A_{\eta_{2}}\left(\alpha U_{2}+\dot{\phi}_{\eta_{2}}\right)\sin\left(\alpha x+\phi_{\eta_{2}}\right)=\nonumber \\
e^{-\alpha\left|z_{1}-z_{2}\right|}A_{w_{1}}\cos\left(\alpha x+\phi_{w_{1}}\right)+A_{w_{2}}\cos\left(\alpha x+\phi_{w_{2}}\right).
\end{eqnarray}
Proceeding in a manner similar to \S \ref{sec:linlin}, the growth-rate $\gamma_{i}$ and phase-speed $c_{i}$ of each wave are found to be
\begin{eqnarray}
 &  & \gamma_{1}=\frac{A_{w_{1}}}{A_{\eta_{1}}}\cos\left(\triangle\phi_{11}\right)+\frac{A_{w_{2}}}{A_{\eta_{1}}}e^{-\alpha\left|z_{1}-z_{2}\right|}\cos\left(\triangle\phi_{12}\right) \label{eq:23}\\
 &  & c_{1}=U_{1}-\frac{1}{\alpha}\left[\frac{A_{w_{1}}}{A_{\eta_{1}}}\sin\left(\triangle\phi_{11}\right)+\frac{A_{w_{2}}}{A_{\eta_{1}}}e^{-\alpha\left|z_{1}-z_{2}\right|}\sin\left(\triangle\phi_{12}\right)\right] \label{eq:24}\\
 &  & \gamma_{2}=\frac{A_{w_{2}}}{A_{\eta_{2}}}\cos\left(\triangle\phi_{22}\right)+\frac{A_{w_{1}}}{A_{\eta_{2}}}e^{-\alpha\left|z_{1}-z_{2}\right|}\cos\left(\triangle\phi_{21}\right) \label{eq:25}\\
 &  & c_{2}=U_{2}-\frac{1}{\alpha}\left[\frac{A_{w_{2}}}{A_{\eta_{2}}}\sin\left(\triangle\phi_{22}\right)+\frac{A_{w_{1}}}{A_{\eta_{2}}}e^{-\alpha\left|z_{1}-z_{2}\right|}\sin\left(\triangle\phi_{21}\right)\right]. \label{eq:26}
\end{eqnarray}
Here  $\triangle\phi_{ij} \equiv \phi_{w_{j}}-\phi_{\eta_{i}}$. %When $i=j$, the angle $\triangle\phi_{ii}$ denotes the phase difference between the wave form and the vertical velocity produced by it. 
When $\alpha\left|z_{1}-z_{2}\right|\rightarrow\infty$, the two waves  decouple, and we recover (\ref{eq:13})-(\ref{eq:14}) for each wave. As argued in \S \ref{sec:linlin}, 
a wave in isolation cannot grow or decay on its own. Therefore, the first term in each of  (\ref{eq:23}) and  (\ref{eq:25}) should be equal to zero, implying  $\left| \triangle \phi_{ii} \right|=\pi/2$. In all our analyses,  we will be considering a system 
with a \emph{left moving top wave} ($\triangle\phi_{11}=\pi/2$) and a \emph{right moving bottom wave} ($\triangle\phi_{22}=-\pi/2$), the wave motion being relative to the background velocity at the corresponding interface. Therefore, we will \emph{only} consider counter-propagating waves (i.e.\ waves moving in a direction opposite to the background flow at that location). \cite{carp2012} provides a detailed explanation how  counter-propagating vorticity waves in Rayleigh's shear layer naturally satisfies Rayleigh-Fj$\o$rtoft's necessary condition of shear instability.

\begin{figure}
\centering
\includegraphics[scale=0.08]{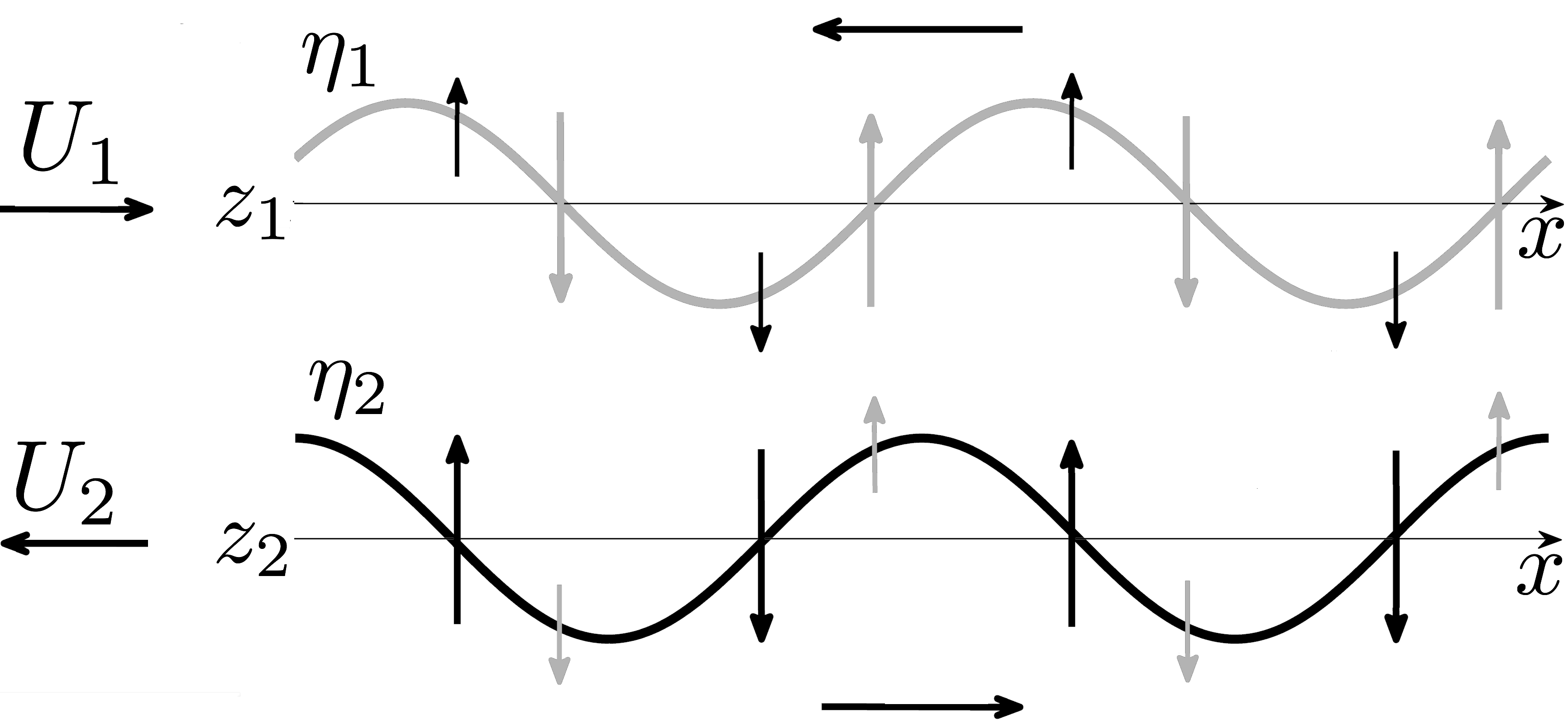}
\setlength{\belowcaptionskip}{-20pt}
\caption{Schematic of the  wave interaction mechanism. The interfacial deformation and associated vertical velocity of each wave are shown by the same colour. Interaction
imposes an additional vertical velocity (shown by a different colour).  The horizontal arrow associated with a wave indicates the intrinsic wave propagation direction. Both waves counter-propagate, i.e.\ move opposite to the background velocity at that location.\bigskip{}}
\label{fig:wave_int}
\end{figure}

Let the phase-shift between the bottom and top waves be
$\Phi \equiv \phi_{\eta_{2}}-\phi_{\eta_{1}}$. Therefore $\triangle\phi_{12}=\Phi-\pi/2$ and $\triangle\phi_{21}=\pi/2-\Phi$. Defining amplitude-ratio $R  \equiv A_{\eta_{1}}/A_{\eta_{2}}$, we re-write (\ref{eq:23})-(\ref{eq:26}) to obtain
\begin{eqnarray}
 &  & \gamma_{1}=\frac{\upomega_{2}}{R} e^{-\alpha\left|z_{1}-z_{2}\right|}\sin\Phi \label{eq:27}\\
 &  & c_{1}=U_{1}-\frac{1}{\alpha}\left[\upomega_{1}-\frac{\upomega_{2}}{R}e^{-\alpha\left|z_{1}-z_{2}\right|}\cos\Phi\right]\label{eq:28}\\
 &  & \gamma_{2}=R\upomega_{1}e^{-\alpha\left|z_{1}-z_{2}\right|}\sin\Phi\label{eq:29}\\
 &  & c_{2}=U_{2}+\frac{1}{\alpha}\left[\upomega_{2}-R\upomega_{1}e^{-\alpha\left|z_{1}-z_{2}\right|}\cos\Phi\right].\label{eq:30}
 \end{eqnarray}
The quantity $\upomega_{i} \equiv A_{w_{i}}/A_{\eta_{i}}$, redefined here for convenience, is constant according to the argument in \S \ref{sec:linlin}.  Eqs.\ (\ref{eq:27})-(\ref{eq:30}) describe the linear hydrodynamic stability of the system. Unlike the conventional linear stability analysis,  we did not impose normal-mode type perturbations (they
only account for exponentially growing instabilities) in our derivation. 
 Therefore the equation-set  provides  a \emph{non-modal}  
 description of hydrodynamic stability in multi-layered  shear flows.  We refer to this theory as the ``Wave Interaction Theory (WIT)''. %WIT is only applicable to those hydrodynamic stability problems where the discrete
 %spectrum dynamics is of interest and the continuous spectrum can be neglected.
 A schematic description of the wave interaction process is illustrated in figure \ref{fig:wave_int}.

Interestingly, there exists an analogy between WIT and the theory behind  \emph{synchronization} of two coupled harmonic oscillators. Synchronization is the process by which interacting, oscillating
objects affect each other's phases such that they spontaneously lock to a certain frequency or phase \cite[]{PIK01}. Weakly (linearly) coupled oscillators interact only through their phases, however more
complicated interaction takes place when the amplitudes of oscillation cannot be neglected. The first analytical step to include the effect of amplitude in the synchronization problem is 
by assuming the coupling to be weakly non-linear. For the case of two coupled harmonic oscillators, weakly non-linear theory yields a system of $4$ equations  \cite[Equation (8.13)]{PIK01},
which represents a dynamical system describing the temporal variation of amplitude and phase  of each oscillator (i.e.\ $\dot{A}_{\eta_{i}}$  and $\dot{\phi}_{\eta_{i}}$). Once the highest order terms are neglected, these equations become analogous to our WIT equation-set (\ref{eq:27})-(\ref{eq:30}). Notice that the WIT equation-set is expressed in terms of $\gamma_{i}$ and $c_{i}$, hence in order to see the correspondence between the two sets, we recall the following relations: $\gamma_{i} \equiv \dot{A}_{\eta_{i}}/A_{\eta_{i}}$ and  $c_{i}\equiv -\dot{\phi}_{\eta_{i}}/\alpha$.

 Observing the analogy with the theory of coupled oscillators, we re-frame the wave interaction problem into a dynamical systems problem.  Subtracting  (\ref{eq:29}) from  (\ref{eq:27}) and  (\ref{eq:30}) from   (\ref{eq:28}), we  find
 \begin{eqnarray}
 &  & \dot{R} =R\left(\gamma_{1}-\gamma_{2}\right)=\left(\upomega_{2}-R^{2}\upomega_{1}\right)e^{-\alpha\left|z_{1}-z_{2}\right|}\sin\Phi \,\,\,\,\,\,\,\,\,\,\, \label{eq:dRdt}\\
 &  & \dot{\Phi}=\alpha\left( c_{1}-c_{2}\right)=\alpha\left(U_{1}-U_{2}\right)-\left[\upomega_{1}+\upomega_{2}-\left(R\upomega_{1}+\frac{\upomega_{2}}{R}\right)e^{-\alpha\left|z_{1}-z_{2}\right|}\cos\Phi\right]. \,\,\,\,\,\,\,\,\,\,\, \label{eq:dPhidt}
\end{eqnarray}
The two parameters have the following range of values: $R\in(0,\,\infty)$ and $\Phi\in[-\pi,\,\pi]$.  
Eq.\ (\ref{eq:dPhidt}) resembles the well known Adler's equation in synchronization theory  \cite[Equation (7.29)]{PIK01}.  The equation-set (\ref{eq:dRdt})-(\ref{eq:dPhidt}) represents a two dimensional, autonomous, non-linear dynamical system. Although the dynamical system is non-linear, the fact that it has only two dynamical variables ($R$ and $\Phi$) imply that the system is \emph{non-chaotic}.  The \emph{two} equilibrium points
of the system,   found by imposing the steady state condition $\dot{R}=0$ in  (\ref{eq:dRdt}) and $\dot{\Phi}=0$ in  (\ref{eq:dPhidt}), are  given by
 \begin{equation}
 (R,\,\Phi)=(R_{NM},\Phi_{NM})\,\,\,\, \mathrm{and}\,\,\,\, (R_{NM},-\Phi_{NM}),
 \label{eq:normode}
 \end{equation}
 where%\footnote{Subscript NM means ``normal-mode'', which will be evident in \S \ref{sec:matstuff}.} 
\begin{eqnarray}
 &  & R_{NM}=\sqrt{\frac{\upomega_{2}}{\upomega_{1}}} \label{eq:R_nm_general}\\
 &  & \Phi_{NM}= \cos^{-1}\left[\left\{ \frac{\upomega_{1}+\upomega_{2}-\alpha\left(U_{1}-U_{2}\right)}{2\sqrt{\upomega_{1}\upomega_{2}}}\right\} e^{\alpha\left|z_{1}-z_{2}\right|}\right]. \label{eq:theta_nm_general}
\end{eqnarray}

For reasons that will become evident in \S \ref{sec:matstuff} we have used the subscript NM, which is short for  ``normal-mode''.
 Eq.\ (\ref{eq:theta_nm_general}) reveals that the equilibrium points exist only if
 \begin{equation}
\left|\left\{ \frac{\upomega_{1}+\upomega_{2}-\alpha\left(U_{1}-U_{2}\right)}{2\sqrt{\upomega_{1}\upomega_{2}}}\right\} e^{\alpha\left|z_{1}-z_{2}\right|} \right|\leq 1.
 \label{eq:necessary}
 \end{equation}
 It might appear that $\Phi=n\pi$, where $n=\{-1, 0, 1 \}$, trivially satisfies the steady state condition. Although (\ref{eq:dRdt}) is satisfied, substitution of $\Phi=n\pi$ in (\ref{eq:dPhidt}) yields only one valid expression of $R$, which is (\ref{eq:R_nm_general}). Any other expression 
  violates (\ref{eq:necessary}) and yields imaginary solutions of $R$ for all $\alpha>0$. 

 The linear behaviour of the dynamical system around the equilibrium points is of interest. To understand this behaviour, we evaluate the Jacobian matrix, $\mathcal{J}$, at the equilibrium points:
 \begin{equation}
\mathcal{J}\left(R_{NM},\pm\Phi_{NM}\right)=-2\sqrt{\upomega_{1}\upomega_{2}}e^{-\alpha\left|z_{1}-z_{2}\right|}\left[\begin{array}{cc}
\sin\left(\pm\Phi_{NM}\right) & 0\\
0 & \sin\left(\pm\Phi_{NM}\right)
\end{array}\right].
\label{eq:jacobian}
\end{equation}
Eq.\ (\ref{eq:jacobian})  shows that the two eigenvalues corresponding to each equilibrium point are equal. Further analysis reveals that every vector at the equilibrium
point is  an eigenvector. The equilibrium point $(R_{NM},\Phi_{NM})$ produces negative eigenvalues (assuming $0 \le \Phi_{NM} \le \pi$), while the eigenvalues corresponding to $(R_{NM},-\Phi_{NM})$ are positive.  
 The dynamical system represented by  (\ref{eq:dRdt})-(\ref{eq:dPhidt}) is therefore ``source-sink'' type. This means that a point in the phase space will move away from the ``source node'' $(R_{NM},-\Phi_{NM})$, and following a unique trajectory, will finally converge to the ``sink node'' $(R_{NM},\Phi_{NM})$.

In summary,  WIT provides an understanding of hydrodynamic instability from two different perspectives -   wave interaction  and  dynamical systems. According to the former, 
exponentially growing instabilities signify resonant interaction between two waves.  From dynamical systems point of view, resonance  implies  a ``steady state or equilibrium condition'' ($\dot{\Phi}=0$ and $\dot{R}=0$). The wave interaction interpretation of each of the two components of the equilibrium condition is as follows:

\begin{figure}

\centering
\subfloat[]{\includegraphics[scale=0.14]{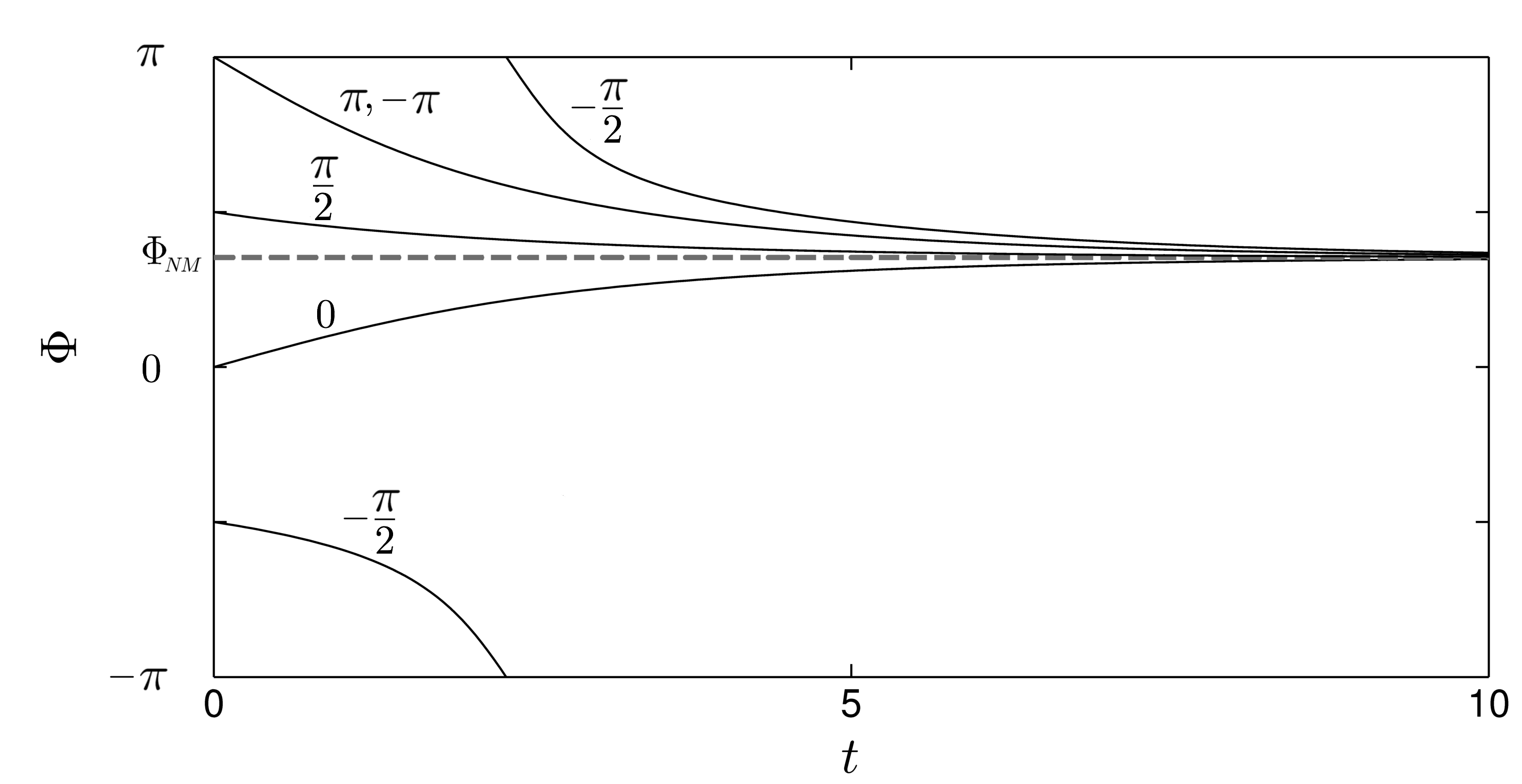}
}

\centering
\subfloat[]{\includegraphics[scale=0.14]{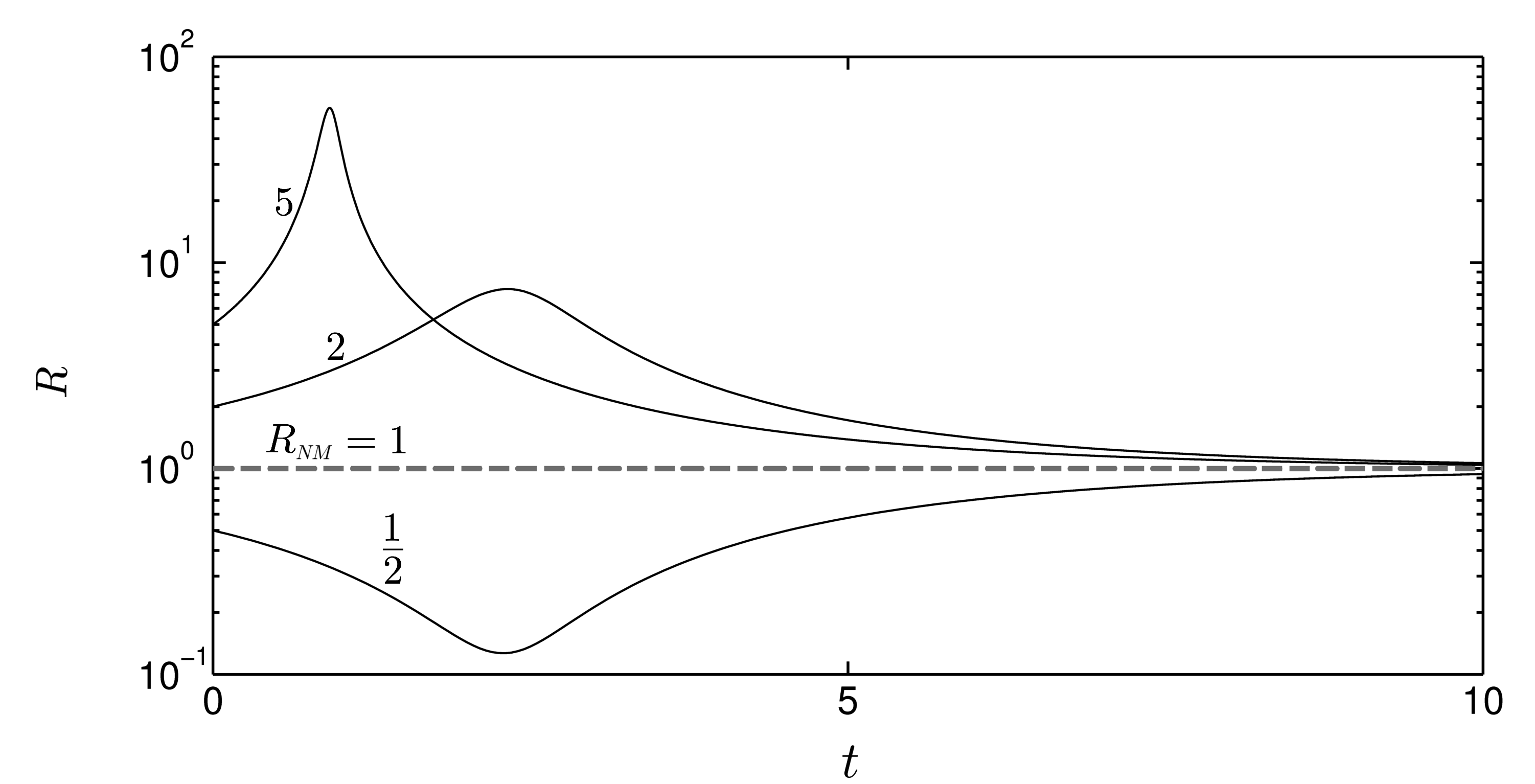}
}
%\centering
%\subfloat[]{\includegraphics[scale=0.085]{t_steady_state}
%}
\setlength{\belowcaptionskip}{-20pt}
\caption{Any initial condition $(R_{0},\Phi_{0})$ finally yields  the resonant configuration $(R_{NM},\Phi_{NM})$, provided   (\ref{eq:necessary}) is satisfied. The case depicted here is  KH instability (interaction between two vorticity waves) corresponding to $\alpha=0.4$. Any other shear instability will show qualitatively similar characteristics. (a) $\Phi$ versus $t$ corresponding to $\Phi_{0} = -\pi, - \pi/2, 0, \Phi_{NM}, \pi/2$ and $\pi$. The value of $R_{0}$ is held constant, and is equal to  $2$. (b) $R$ versus $t$ corresponding to $R_{0} = 1/2, 1 (R_{NM}), 2$ and $5$. The value of $\Phi_{0}$ is held constant, and is equal to  $-\pi/2$.\bigskip{}}
\label{fig:Summaryfig}

\end{figure}

\paragraph{(a)} \textit{Phase-Locking} or $\dot{\Phi}=0$: 
Reduction in the  phase-speed of each wave occurs through the interaction mechanism - the vertical velocity field produced by the distant wave acts so as to diminish the phase-speed of the given wave. Furthermore, if the waves are
``counter-propagating''
(meaning, the direction of  $c_{i}^{intr}$ is opposite to the background flow), the background flow causes an additional reduction in the phase-speed. Both wave interaction
and counter-propagation work synergistically until the two waves become ``phase-locked'', i.e.\  stationary relative to each other ($\dot{\Phi}=0$). The phase-shift at the phase-locked state, $\Phi_{NM}$, is a unique angle dictated by the physical parameters of the system, as evident from (\ref{eq:theta_nm_general}).
\emph{Any} arbitrary initial condition (say $R=R_{0}$ and $\Phi=\Phi_{0}$)  finally leads to  phase-locking, as evident from  figure \ref{fig:Summaryfig}(a), provided (\ref{eq:necessary}) is satisfied. When the condition (\ref{eq:necessary}) is \emph{not} satisfied, the interaction is too weak to cause phase locking. The two waves continue moving in opposite directions, hence the 
 phase-shift angle oscillates between $[0,2\pi]$ (for example see figure \ref{fig:manyfigs}(d)).

\paragraph{(b)} \textit{Mutual Growth} or  $\dot{R}=0$: Not only the two waves lock in phase, they also lock in amplitude, producing the unique steady state amplitude-ratio $R_{NM}$; see figure \ref{fig:Summaryfig}(b). Eq.\ (\ref{eq:dRdt}) reveals that   $\dot{R}=0$  implies \ $\gamma_{1}=\gamma_{2}$, which in turn signifies \emph{resonance} between the two waves.  
Furthermore,  (\ref{eq:27}) and (\ref{eq:29}) imply  that at steady state $\gamma_{1}=\gamma_{2}=\mathrm{constant}$, meaning that the wave amplitudes grow at an exponential rate.

\section{Modal and non-modal analysis}
\label{sec:matstuff}

In the previous section we used WIT and dynamical systems theory to describe multi-layered inviscid hydrodynamic instabilities in terms of interacting interfacial waves.  We found that the ``equilibrium condition'' of the dynamical system basically signifies the ``resonant condition'' of WIT. A question which naturally arises is ``\emph{what do these conditions mean in terms of conventional hydrodynamic stability theory}?''  In this section we will draw parallels between WIT (and the dynamical systems formulation) and eigenanalysis and Singular Value Decomposition (SVD), which are conventionally used to study modal and non-modal instabilities, respectively.

%The conventional approach to studying linear shear stability problems is through matrices; normal-mode type instability is studied through eigenanalysis and non-normal instabilities through Singular Value Decomposition (SVD). In the previous section we understood linear hydrodynamic instabilities using non-traditional approaches. First, we proposed WIT, which facilitates understanding hydrodynamic instabilities in multi-layered inviscid flows in terms of interacting interfacial waves. This was done by transforming the linearized kinematic condition with two interfaces, i.e.\ (\ref{eq:kin1})-(\ref{eq:kin2}), to Fourier space. Next, we re-framed the WIT equations into a non-linear, autonomous dynamical system (\ref{eq:dRdt})-(\ref{eq:dPhidt}). Furthermore we found that the ``equilibrium condition'' of the dynamical system basically signifies the ``resonant condition'' of WIT. A question which naturally arises is ``\emph{what do these conditions mean in terms of conventional hydrodynamic stability theory}?'' In this section we will introduce eigenanalysis and SVD, and draw parallels with WIT (and the dynamical systems formulation).

\subsection{Eigenanalysis and SVD}
\label{subsec:matrixtheory}

 The linearized kinematic condition with two interfaces, (\ref{eq:kin1})-(\ref{eq:kin2}),  can be written in the matrix form by expressing $w_{i}$ in terms of $\eta_{i}$ (by invoking the definition of $\upomega_{i}$), and imposing the condition for counter-propagation ($\triangle\phi_{11}=\pi/2$ and $\triangle\phi_{22}=-\pi/2$):
 
\begin{equation}
{\frac{\partial \boldsymbol{\eta}}{\partial t}=\boldsymbol{\mathcal{M}\eta}},
\label{eq:eettaa}
\end{equation}

where

\begin{equation}
\boldsymbol{\eta}=\left[\begin{array}{c}
\eta_{1}\\
\eta_{2}
\end{array}\right]\,\,\,\,\,\,\,\,\,\textrm{and}\,\,\,\,\,\,\,\,\boldsymbol{\mathcal{M}}=-i\left[\begin{array}{cc}
\alpha U_{1}-\upomega_{1} & \upomega_{2}e^{-\alpha\left|z_{1}-z_{2}\right|}\\
-\upomega_{1}e^{-\alpha\left|z_{1}-z_{2}\right|} & \alpha U_{2}+\upomega_{2}
\end{array}\right].
\end{equation}

%\begin{equation}
%\boldsymbol{\eta}=\left[\begin{array}{c}
%\eta_{1}\\
%\eta_{2}
%\end{array}\right]
%\end{equation}
%
%
%and
%
%\begin{equation}
%\boldsymbol{\mathcal{M}}=-\ii\left[\begin{array}{cc}
%\alpha U_{1}-\upomega_{1} & \upomega_{2}e^{-\alpha\left|z_{1}-z_{2}\right|}\\
%-\upomega_{1}e^{-\alpha\left|z_{1}-z_{2}\right|} & \alpha U_{2}+\upomega_{2}
%\end{array}\right].
%\end{equation}
 Eq.\ (\ref{eq:eettaa}) represents the first order perturbation dynamics, and  $\boldsymbol{\mathcal{M}}$ is the linearized dynamical operator. Since our dynamical system is autonomous $\boldsymbol{\mathcal{M}}$ is time independent, and the solution is explicit:
\begin{equation}
\boldsymbol{\eta}\left(t\right)=e^{\boldsymbol{\mathcal{M}}t}\boldsymbol{\eta}\left(0\right)=\left[\boldsymbol{P}e^{\boldsymbol{L}t}\boldsymbol{P}^{-1}\right]\boldsymbol{\eta}\left(0\right)=\left[\boldsymbol{\mathcal{U}}\boldsymbol{\Sigma}\boldsymbol{V}^{\dagger}\right]\boldsymbol{\eta}\left(0\right).
\label{eq:boroeqn}
\end{equation}
The matrix exponential $e^{\boldsymbol{\mathcal{M}}t}$ in the above equation is the propagator matrix, which advances the system in time. The transient dynamics of the system is solely governed by the normality of  $\boldsymbol{\mathcal{M}}$,  i.e.\ whether or not $\boldsymbol{\mathcal{M}}$ commutes with its Hermitian transpose ${\boldsymbol{\mathcal{M}}^{\dagger}}$  \cite[]{farrell1996}. If they commute ($\boldsymbol{\mathcal{M}}\boldsymbol{\mathcal{M}}^{\dagger}=\boldsymbol{\mathcal{M}}^{\dagger}\boldsymbol{\mathcal{M}}$) then  $\boldsymbol{\mathcal{M}}$  is normal and has complete set of orthogonal eigenvectors. In this case the dynamics can be fully understood from the eigendecomposition of the propagator. This basically means expressing $e^{\boldsymbol{\mathcal{M}}t}$ as $\boldsymbol{P}e^{\boldsymbol{L}t}\boldsymbol{P}^{-1}$, where  $\boldsymbol{L}$ is a diagonal matrix containing the complex eigenvalues $\lambda$ of $\boldsymbol{\mathcal{M}}$ (arranged by the real part of the eigenvalues in the descending order of magnitude), and $\boldsymbol{P}$ is the corresponding matrix of eigenvectors. Alternatively if $\boldsymbol{\mathcal{M}}$ is non-normal, the interaction between the discrete non-orthogonal modes
of $\boldsymbol{\mathcal{M}}$ produces non-normal growth processes. Non-normality can be understood through SVD of the propagator. SVD is a generalized matrix factorization technique and matches eigendecomposition \emph{only} when $\boldsymbol{\mathcal{M}}$ is Hermitian ($\boldsymbol{\mathcal{M}}=\boldsymbol{\mathcal{M}}^{\dagger}$). SVD of the propagator yields   $\boldsymbol{\mathcal{U}}\boldsymbol{\Sigma}\boldsymbol{V}^{\dagger}$, where $\boldsymbol{\mathcal{U}}$
contains the eigenvectors of $e^{\boldsymbol{\mathcal{M}}t}e^{\boldsymbol{\mathcal{M}}^{\dagger}t}$, $\boldsymbol{V}$ contains the eigenvectors of $e^{\boldsymbol{\mathcal{M}}^{\dagger}t}e^{\boldsymbol{\mathcal{M}}t}$,
and $\boldsymbol{\Sigma}$ is a diagonal matrix containing the singular values ($\sigma$, which are real and positive) arranged in the descending order of magnitude. If  $\boldsymbol{\mathcal{M}}$ commutes with its Hermitian transpose, then $\sigma_{max}=e^{\Re (\lambda_{max}) t}$. Otherwise $\sigma_{max}>e^{\Re (\lambda_{max}) t}$, implying that growth-rate higher than the least stable normal-mode is possible \cite[]{farrell1996}.

 \subsubsection{Eigenanalysis} 
\label{subsub:eigenanalysis}
If $\boldsymbol{\mathcal{M}}$ is a normal matrix, then eigendecomposition of the propagator matrix is sufficient to capture the dynamics. If $\boldsymbol{\mathcal{M}}$ is  non-normal, then eigenanalysis captures the asymptotic dynamics  \emph{only} for large times. 

The eigenvalues of the matrix $\boldsymbol{\mathcal{M}}$ are as follows:
\begin{equation}
\lambda_{\pm}=-\frac{\ii}{2}\left[\alpha\left(U_{1}+U_{2}\right)-\left(\upomega_{1}-\upomega_{2}\right)\right]\pm\frac{1}{2}\sqrt{\mathscr{D}},
\label{eq:lamd}
\end{equation}
where $\mathscr{D}=4\upomega_{1}\upomega_{2}e^{-2\alpha\left|z_{1}-z_{2}\right|}-\left[\alpha\left(U_{1}-U_{2}\right)-\left(\upomega_{1}+\upomega_{2}\right)\right]^{2}$. Normal-mode instability can \emph{only} occur if $\mathscr{D}>0$. This basically gives rise to the condition (\ref{eq:necessary}), implying that (\ref{eq:necessary}) denotes the \emph{necessary and sufficient} (N\&S) \emph{condition} for exponentially
growing instabilities in idealized (broken-line profiles), homogeneous and stratified, inviscid shear layers. The \emph{unstable} eigenvalues of (\ref{eq:lamd}) can therefore be written as
\begin{equation}
\lambda_{\pm}=-\frac{\ii}{2}\left[\alpha\left(U_{1}+U_{2}\right)-\left(\upomega_{1}-\upomega_{2}\right)\right]\pm\sqrt{\upomega_{1}\upomega_{2}}e^{-\alpha\left|z_{1}-z_{2}\right|}\sin\left(\Phi_{NM}\right).
\label{eq:lamd1}
\end{equation}
The positive and negative eigenvalues respectively imply growing and decaying normal-modes of the discrete spectrum. There is a direct relation between the normal-modes and the resonant condition of WIT (or the equilibrium condition of the dynamical system (\ref{eq:dRdt})-(\ref{eq:dPhidt})). \emph{Each equilibrium point corresponds to a normal-mode} :   $(R_{NM},\Phi_{NM})$ corresponds to the growing normal-mode 
(signifying exponential growth $e^{ \Re (\lambda_{+})t}$), and $(R_{NM},-\Phi_{NM})$ corresponds to the decaying normal-mode (signifying exponential decay $e^{ \Re (\lambda_{-})t}$). An analogy can be drawn with a system of two coupled harmonic oscillators. Such a system has two normal-modes of vibration, the in-phase synchronization mode (zero phase-shift)  and the anti-phase synchronization mode (phase-shift of $\pi$). The in-phase and anti-phase modes are respectively analogous to the growing and decaying normal-modes of WIT.    

WIT is only relevant to problems where the discrete spectrum is of primary interest, and the effect of the continuous spectrum can be neglected. In other words, the applicability of WIT is restricted to idealized multi-layered profiles. Real profiles are continuous and their eigenanalysis yields a continuous spectrum. Although the continuous spectrum consists of neutral modes, they interact to produce sheared disturbances. Such disturbances can be obtained by artificially changing the vorticities of fluid elements by small amounts, imposing a sinusoidal streamwise dependence, and then letting the system  evolve freely. This spectrum crucially contributes to the non-modal instability processes, and is responsible for large transient growths.

 \subsubsection{SVD analysis} 
 If $\boldsymbol{\mathcal{M}}$ is  non-normal, then  eigenanalysis fails to capture the dynamics in the limit $t\rightarrow 0$.  To understand the short-term dynamics, one needs to perform SVD analysis. The eigenvalue matrix of  $e^{\boldsymbol{\mathcal{M}}t}e^{\boldsymbol{\mathcal{M}}^{\dagger}t}$ (or  $e^{\boldsymbol{\mathcal{M}}^{\dagger}t}e^{\boldsymbol{\mathcal{M}}t}$) is equal to $\boldsymbol{\Sigma}^{2}$; the maximum eigenvalue of $\boldsymbol{\Sigma}$ is key in determining the transient dynamics.  In the limit $t\rightarrow 0$, Taylor expansion of $e^{\boldsymbol{\mathcal{M}}^{\dagger}t}e^{\boldsymbol{\mathcal{M}}t}$ produces \cite[]{farrell1996}:
\begin{eqnarray}
\ensuremath{e^{\boldsymbol{\mathcal{M}}^{\dagger}t}e^{\boldsymbol{\mathcal{M}}t} \approx ( \boldsymbol{I}+\boldsymbol{\mathcal{M}}^{\dagger}t+ \ldots ) (\boldsymbol{I}+\boldsymbol{\mathcal{M}}t+\ldots )} \nonumber \\
=\boldsymbol{I} + ( \boldsymbol{\mathcal{M}}+\boldsymbol{\mathcal{M}}^{\dagger}) t+O (t^{2}),\,\,\,\,\,\,\,\,\,\,\,\,\,\,\,\,
\end{eqnarray}
where  $\boldsymbol{I}$ is the identity matrix. The maximum eigenvalue of $\frac{1}{2} (\boldsymbol{\mathcal{M}}+\boldsymbol{\mathcal{M}}^{\dagger})$ (which is known as the \emph{numerical abscissa} of $\boldsymbol{\mathcal{M}}$)  and its associated eigenvector provide the maximum instantaneous growth-rate and structure. The numerical abscissa is found to be
\begin{equation}
\sigma_{max}=\frac{1}{2}\left(\upomega_{1}+\upomega_{2}\right)e^{-\alpha\left|z_{1}-z_{2}\right|}.
\end{equation}
It is straight-forward to check that
\begin{equation}
\frac{\sigma_{max}}{\Re\left(\lambda_{+}\right)}\geq\frac{1}{\sin\left(\Phi_{NM}\right)}\Rightarrow\sigma_{max}\geq\Re\left(\lambda_{+}\right).
\end{equation}

\subsection{Non-modal  growth}
\label{subsec:non_mod}

The non-normality of the system can give rise to transient energy amplification. Even when the waves are decaying, the non-orthogonal
superposition of eigenvectors may lead to short-time growth of energy (or some other norm). For the non-modal analysis we have chosen % the mean squared wave amplitude, %$E=\frac{1}{2}\left(A_{\eta_{1}}^{2}+A_{\eta_{2}}^{2}\right)$
$E=\upomega_{1}A_{\eta_{1}}^{2}+\upomega_{2}A_{\eta_{2}}^{2}$ as a suitable norm. %This choice is based on the WIT formulation.%; alternative norms like  $A_{\eta_{1}}^{2}+ A_{\eta_{2}}^{2}$ or $\upomega_{1}^{2} A_{\eta_{1}}^{2}+\upomega_{2}^{2}A_{\eta_{2}}^{2}$ yield  less straight-forward  analysis in \S \ref{subsub:NSNS}. 
 The time evolution of $E$ is used to indicate the growth (or decay) of the system, and is calculated as follows:
\begin{equation}
 \dot{E}=2\upomega_{1}A_{\eta_{1}}\dot{A}_{\eta_{1}}+2\upomega_{2}A_{\eta_{2}}\dot{A}_{\eta_{2}}=4A_{\eta_{1}}A_{\eta_{2}}\upomega_{1}\upomega_{2}e^{-\alpha\left|z_{1}-z_{2}\right|}\sin\left(\Phi\right).%2A_{\eta_{1}}A_{\eta_{2}}\sigma_{max}\sin\left(\Phi\right).
\label{eq:mejo}
\end{equation}
The rightmost side of (\ref{eq:mejo}) is obtained by  substituting (\ref{eq:27}) and (\ref{eq:29}). 
Furthermore, the inequality $ \left(\sqrt{\upomega_{1}}A_{\eta_{1}} - \sqrt{\upomega_{2}}A_{\eta_{2}}\right)^{2 } \geq 0 \Rightarrow \upomega_{1} A_{\eta_{1}}^{2}+\upomega_{2}A_{\eta_{2}}^{2} \geq 2 \sqrt{\upomega_{1}\upomega_{2}}A_{\eta_{1}}A_{\eta_{2}}$.  Hence
\begin{equation}
%\frac{dE}{dt}  \le  \left(A_{\eta_{1}}^{2}+A_{\eta_{2}}^{2}\right)\sigma_{max}\sin\left(\Phi\right).
 \dot{E} \le 2E \sqrt{\upomega_{1}\upomega_{2}}e^{-\alpha\left|z_{1}-z_{2}\right|}\sin\left(\Phi\right).
\label{eq:jjj1}
\end{equation}
$\dot{E}$ is maximum  when $\left(\sqrt{\upomega_{1}}A_{\eta_{1}} - \sqrt{\upomega_{2}}A_{\eta_{2}}\right)^{2 } = 0 \Rightarrow R=\sqrt{\upomega_{2}/\upomega_{1}}$. The corresponding $E$ is $E_{m}=2\upomega_{1}A_{\eta_{1}}^{2}=2\upomega_{2}A_{\eta_{2}}^{2}$. The time evolution of $\dot{E}_{m}$  is given by 
  %Thus when $A_{\eta_{1}}=A_{\eta_{2}}=A_{\eta}$ (say), i.e. when $R=1$,
%we have the maximum growth condition. Let $E_{max}=A_{\eta}^{2}$ denote the maximum growth norm. 
 %Under this condition (\ref{eq:jjj1}) can be written as
  \begin{equation}
\dot{E}_{m}  = 2E_{m} \sqrt{\upomega_{1}\upomega_{2}}e^{-\alpha\left|z_{1}-z_{2}\right|}\sin\left(\Phi\right).
%\frac{dE_{max}}{dt} = 2E_{max}\sigma_{max}\sin\left(\Phi\right).
\label{eq:Emax}
\end{equation}
The sign of the R.H.S. of (\ref{eq:Emax}) dictates whether the system is growing (positive sign) or decaying (negative sign). Since only $\sin\left(\Phi\right)$ is a signed quantity,  it means that the instantaneous phase-shift governs the instantaneous growth or decay.  The \emph{largest} instantaneous growth occurs when $\Phi=\pi/2$ (this fact can also be verified from (\ref{eq:27}) and (\ref{eq:29})). Figure \ref{fig:manyfigs}(a) shows an example of how $A_{\eta}$ (=$\sqrt{E_{m}/(2\upomega)}$) varies with time. During the initial period, the wave amplitude decays (since the initial phase-shift $\Phi_{0}<0$) and then grows at a rate faster than that predicted by the normal-mode theory. Eventually, the growth-rate asymptotes to that predicted by normal-mode theory.  Note that the flow becomes non-linear relatively quickly, and the linear theories (e.g.\ WIT and normal-mode) start to deviate from the non-linear theory;  see figure 3 of \cite{guha2012}.  

 The amplification or gain ($G$) of the system is given by:   
  \begin{equation}
G\left(t\right)=\frac{E_{max}\left(t\right)}{E_{max}\left(0\right)}=\exp\left[2 \sqrt{\upomega_{1}\upomega_{2}}e^{-\alpha\left|z_{1}-z_{2}\right|} \intop_{0}^{t}\sin\left(\Phi\right)dt'\right].
\label{eq:GG}
\end{equation}

The quantity $dt$ can be expressed in terms of $d\Phi$ by invoking (\ref{eq:dPhidt}). Substituting $R=\sqrt{\upomega_{2}/\upomega_{1}}$ and integrating, we obtain
%\begin{equation}
%G=\left|\frac{\alpha\left(U_{1}-U_{2}\right)-\left(\upomega_{1}+\upomega_{2}\right)\left(1-e^{-\alpha\left|z_{1}-z_{2}\right|}\cos\left(\Phi_{0}\right)\right)}{\alpha\left(U_{1}-U_{2}\right)-\left(\upomega_{1}+\upomega_{2}\right)\left(1-e^{-\alpha\left|z_{1}-z_{2}\right|}\cos\left(\Phi_{t}\right)\right)}\right|.
%\end{equation}
\begin{equation}
%G=\left|\frac{Q+\cos\left(\Phi_{0}\right)}{Q+\cos\left(\Phi_{t}\right)}\right|.
G\left(t\right)=\left|\frac{\cos\left(\Phi_{0}\right)-Q}{\cos\left(\Phi_{t}\right)-Q}\right|,
%G=\left|\frac{Q+\cos\left(\Phi_{0}\right)}{\frac{Q+\cos\left(\Phi_{t}\right)}\right|,
\label{eq:GG1}
\end{equation}
where $Q=\{\upomega_{1}+\upomega_{2}-\alpha(U_{1}-U_{2})\}e^{\alpha\left|z_{1}-z_{2}\right|}/(2 \sqrt{\upomega_{1}\upomega_{2}})$, and $\Phi_{t}$ is the value of $\Phi$ at $t=t$. Notice that $|Q| \leq 1$ denotes the N\&S condition (\ref{eq:necessary}). %The two  alternative norms stated previously produce the same expression  (\ref{eq:GG1}), however the value of $Q$ is different.
%where $Q=\left\{ \alpha (U_{1}-U_{2})/(\upomega_{1}+\upomega_{2})-1\right\} e^{\alpha\left|z_{1}-z_{2}\right|}$.

\subsubsection{Non-modal growth when N\&S condition is satisfied:} 
\label{subsub:NSNS}
\begin{figure}

\centering
\subfloat[]{\includegraphics[scale=0.07]{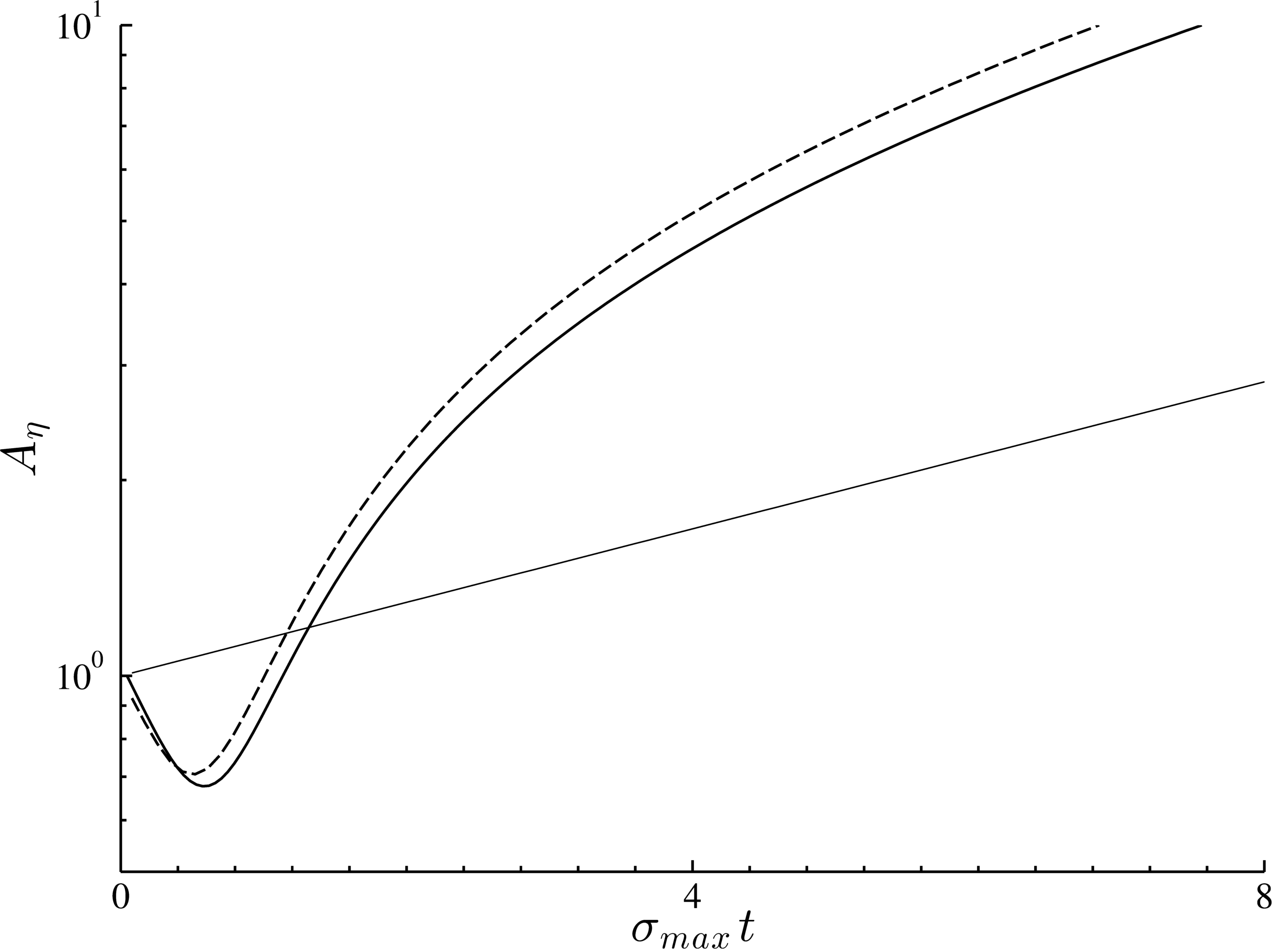}
}
\centering
~~\subfloat[]{\includegraphics[scale=0.095]{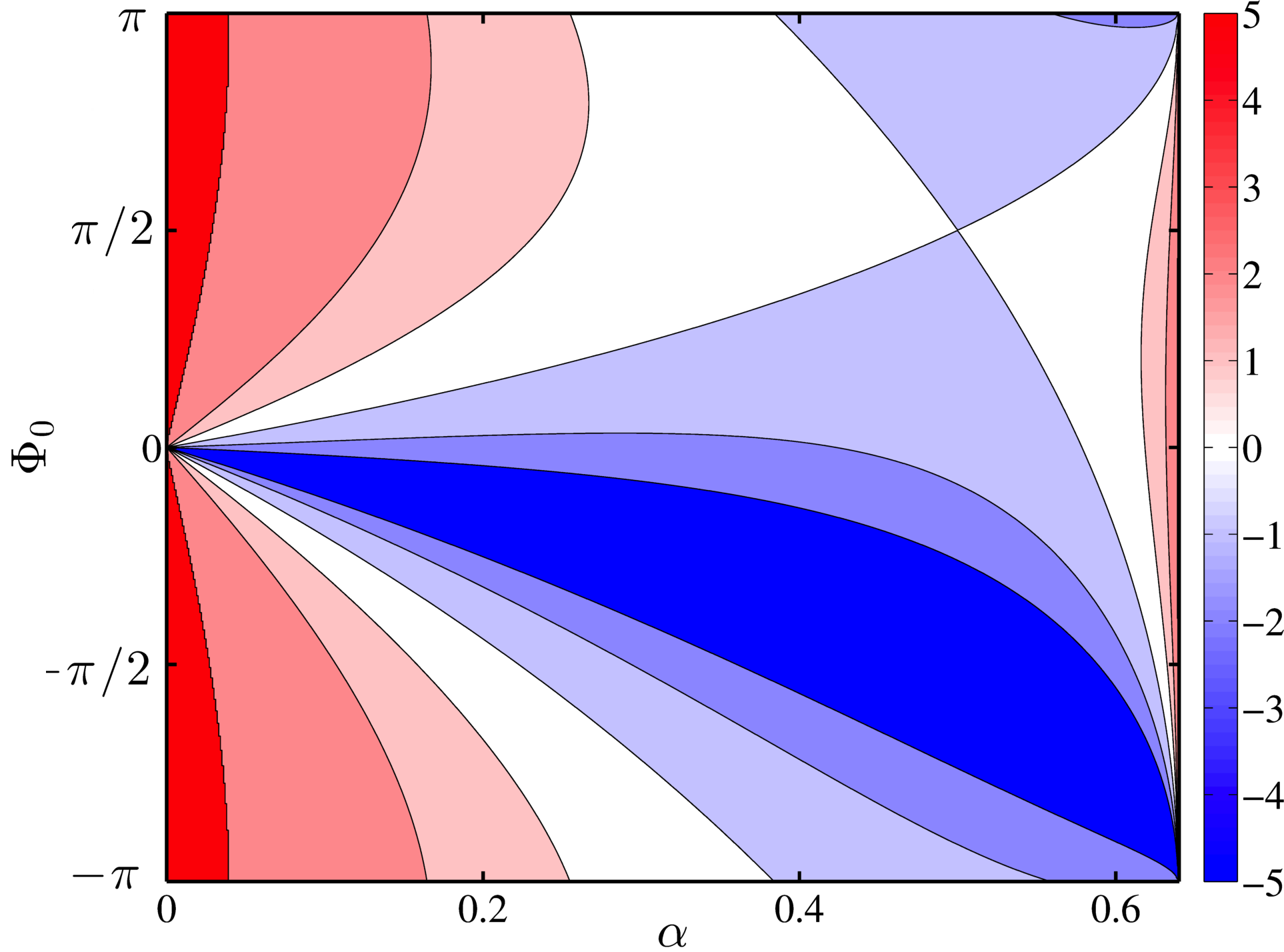}
}

\centering
\subfloat[]{\includegraphics[scale=0.075]{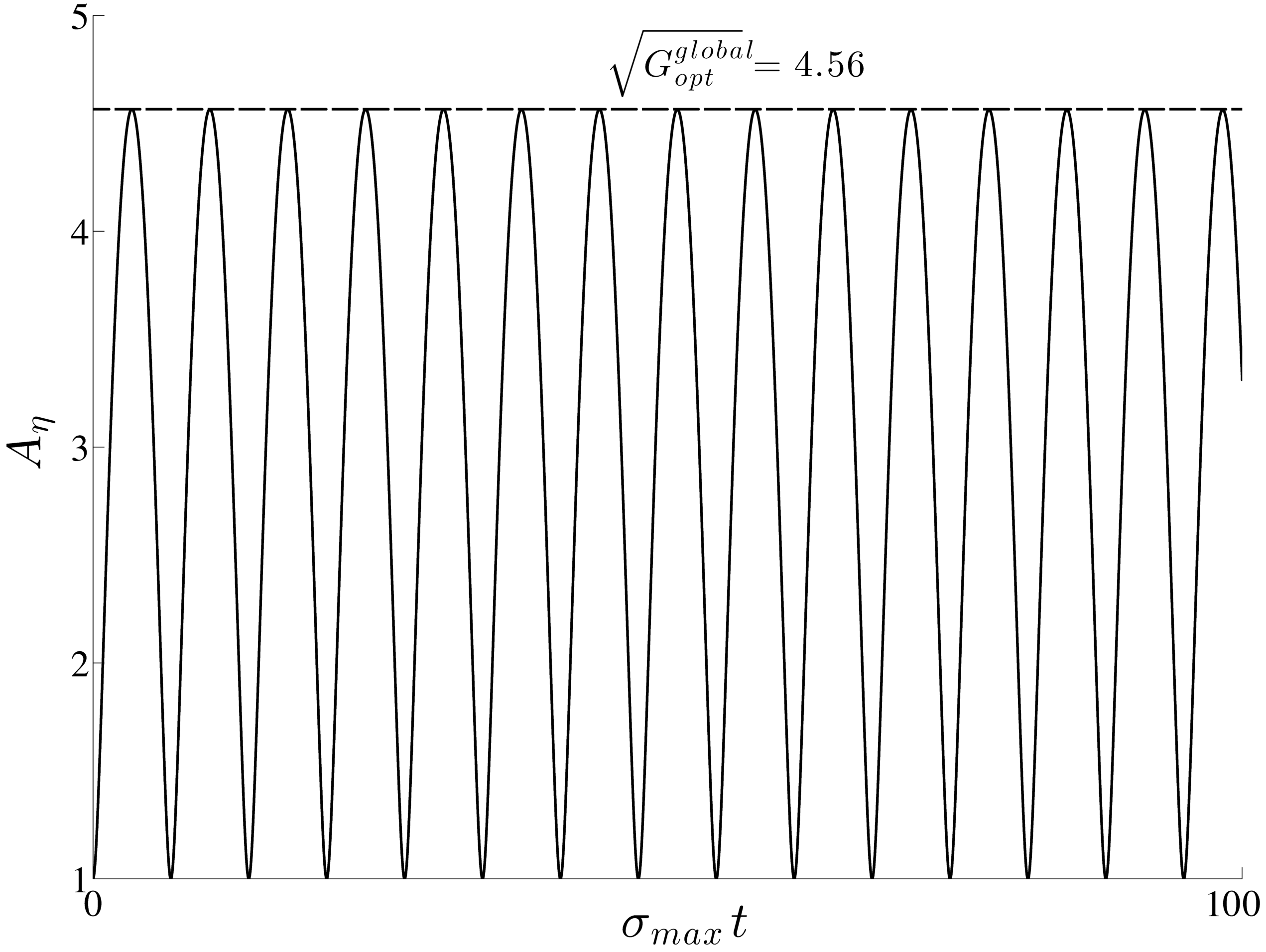}
}
\centering
~~~~~\subfloat[]{\includegraphics[scale=0.075]{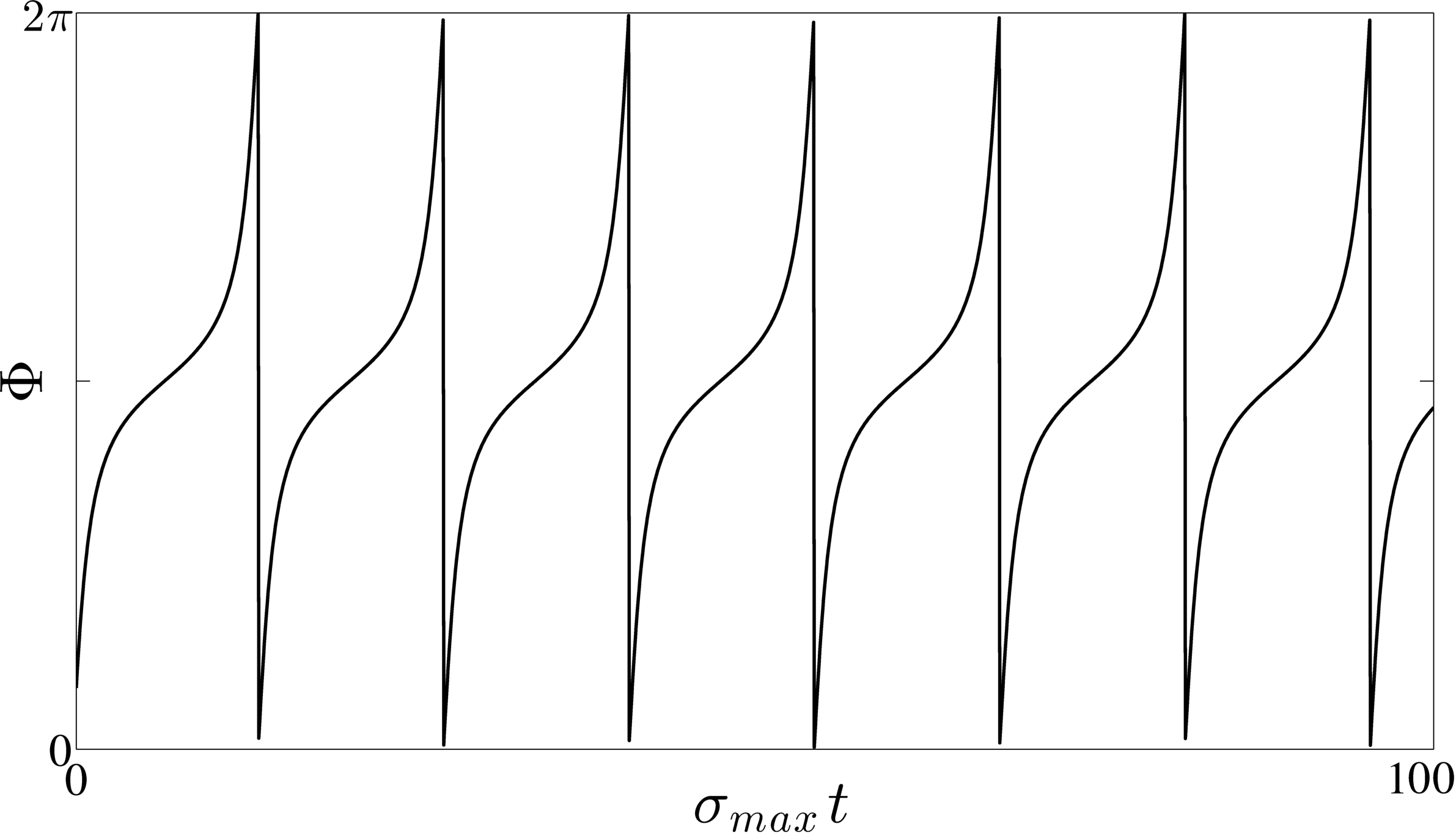}
}
\setlength{\belowcaptionskip}{-10pt}
 \caption{ Different aspects of transient dynamics of KH instability: (a) Variation of  wave amplitude with target time (both normalized) for $\alpha=0.0625$ and $\Phi_{0}=-\pi/2$. The  bold solid curve represents WIT, and the bold dashed curve is its non-linear extension; see \cite{guha2012}. The thin solid line represents amplitude variation predicted by the normal-mode theory. The slope of the WIT curve becomes parallel to that of the normal-mode for large times. (b) (Colour online) Plot of $\log(\chi)$ as a function of $\alpha$ and $\Phi_{0}$. Positive contour values indicate non-modal growth exceeding corresponding modal growth.
 (c) Variation of normalized wave amplitude with normalized target time for $\alpha=0.65$. (d) Variation of phase-shift with normalized target time for $\alpha=0.65$.  %Variation of  $G_{opt}^{global}$ with $\alpha$. The grey region depicts the range of wavenumbers for which the flow is linearly unstable to normal-mode disturbances.  
  \bigskip{}
 }
\label{fig:manyfigs}
\end{figure}

When the N\&S condition is satisfied  $Q=\cos(\Phi_{NM})$; see (\ref{eq:theta_nm_general}). Hence (\ref{eq:GG1}) yields
\begin{equation}
G\left(t\right)=\left|\frac{\sin\left(\frac{\Phi_{NM}+\Phi_{0}}{2}\right)\sin\left(\frac{\Phi_{NM}-\Phi_{0}}{2}\right)}{\sin\left(\frac{\Phi_{NM}+\Phi_{t}}{2}\right)\sin\left(\frac{\Phi_{NM}-\Phi_{t}}{2}\right)}\right|.
\label{eq:GG3}
\end{equation}
 We find that  $G \rightarrow \infty$ when  the final phase-shift $\Phi_{t} \rightarrow \pm\Phi_{NM}$. The positive and negative values of $\Phi_{NM}$ respectively denote the sink node and the source node in  phase space. Since trajectories move away from the source node,  $\Phi_{t} \nrightarrow -\Phi_{NM}$ during the non-modal evolution process (for example, see figure \ref{fig:KH}(b)). 
 The significance of the non-modal gain $G$ can be properly understood when compared  with a hypothetical case in which the system undergoes  modal growth between $\Phi=\Phi_{0}$ and $\Phi_{NM}$.  For this we first compute the gain obtained from a normal-mode type perturbation growing over an interval $t$: 
 
%The importance of the non-modal gain $G$ by comparing it with the modal gain $G_{NM}$.
 %compares to the gain that would be obtained if a perturbation grows exponentially 
 %obtained from a  normal-mode type perturbation. The latter is computed as follows:

\begin{eqnarray}
G_{NM}\left(t\right)=e^{2\Re\left(\lambda_{+}\right)t}=\exp\left[2\left\{ \tanh^{-1}\left(\frac{\tan\left(\frac{\Phi_{t}}{2}\right)}{\tan\left(\frac{\Phi_{NM}}{2}\right)}\right)-\tanh^{-1}\left(\frac{\tan\left(\frac{\Phi_{0}}{2}\right)}{\tan\left(\frac{\Phi_{NM}}{2}\right)}\right)\right\} \right] \nonumber \\
=\left|\frac{\sin\left(\frac{\Phi_{NM}+\Phi_{t}}{2}\right)}{\sin\left(\frac{\Phi_{NM}-\Phi_{t}}{2}\right)}\right|\left|\frac{\sin\left(\frac{\Phi_{NM}-\Phi_{0}}{2}\right)}{\sin\left(\frac{\Phi_{NM}+\Phi_{0}}{2}\right)}\right|.\,\,\,\,\,\,\,\,\,\,\,\,\,\,\,\,\,\,\,\,\,\,\,\,\,\,\,\,\,\,\,\,\,\,\,\,\,\,\,\,\,\,\,\,\,\,\,\,\,\,\,\,\,\,\,\,\,\,\,\,\,\,\,\,\,\,\,\,\,\,
\end{eqnarray}
Like $G$, the normal-mode gain $G_{NM} \rightarrow \infty$ when $\Phi_{t} \rightarrow \Phi_{NM}$. The gain-ratio $\chi\left(t\right)=G(t)/G_{NM}(t)$ is found to be
\begin{equation}
\chi\left(t\right)=\frac{\sin^{2}\left(\frac{\Phi_{NM}+\Phi_{0}}{2}\right)}{\sin^{2}\left(\frac{\Phi_{NM}+\Phi_{t}}{2}\right)}.
\label{eq:GG8}
\end{equation}
When $\chi>1$, the non-normal growth exceeds exponential growth within the time window $[0,t]$. For our calculations we have chosen the entire time window of non-modal evolution. Thus (\ref{eq:GG8}) becomes
\begin{equation}
\chi\left(t_{NM}\right)=\frac{\sin^{2}\left(\frac{\Phi_{NM}+\Phi_{0}}{2}\right)}{\sin^{2}\left(\Phi_{NM}\right)}.
\label{eq:GG18}
\end{equation}
 Notice that unlike $G$ and $G_{NM}$, the value of $\chi$ is finite as $t \rightarrow t_{NM}$. Figure \ref{fig:manyfigs}(b) depicts the variation of $\log(\chi)$ with $\alpha$ and $\Phi_{0}$ for 
KH instability. The flow is linearly unstable for $0 \leq \alpha \leq 0.64$ (see \S \ref{sec:KH}).
 Non-modal gain \emph{exceeds} modal gain by \emph{several} orders of magnitude in the neighbourhood of stability boundaries. This behaviour is especially prominent near the lower boundary.  The fact that substantial transient growth occurs near stability boundaries has also been observed for non-modal Holmboe instability \cite[]{const2011}. The magnitude of $\chi$ is strongly dependent on the initial condition. We observe  higher  values of $\chi$ when the component waves are long (smaller values of $\alpha$), and initially out of phase ($\Phi_{0}$ away from $0$). This situation reverses near the upper boundary; waves starting in-phase exhibit higher magnitudes of $\chi$.

\subsubsection{Non-modal growth when N\&S condition is not satisfied:}
When the N\&S condition is not satisfied $|Q| > 1$, and the flow is linearly stable according to the normal-mode theory. The amplitude of a small perturbation is therefore expected to stay constant in this region. 
  However non-modal processes can produce transient growth in the parameter space
 for which the flow is deemed stable by the normal-mode analysis. This can be shown by studying the evolution of the optimal perturbation. Using SVD analysis  \cite{heif2005}  have shown that optimal perturbation evolves such that the final phase-shift $\Phi_{t}=\pi-\Phi_{0}$. The phase-shift is symmetric in time about $\pi/2$, maximizing $\sin\left(\Phi\right)$ in (\ref{eq:GG}). Thus the optimal gain $G_{opt}$ is found to be
\begin{equation}
G_{opt}=\left|\frac{Q+\cos\left(\Phi_{0}\right)}{Q-\cos\left(\Phi_{0}\right)}\right|.
%G_{opt}=\left|\frac{\left\{ \frac{\alpha\left(U_{1}-U_{2}\right)}{\upomega_{1}+\upomega_{2}}-1\right\} e^{\alpha\left|z_{1}-z_{2}\right|}+\cos\left(\Phi_{0}\right)}{\left\{ \frac{\alpha\left(U_{1}-U_{2}\right)}{\upomega_{1}+\upomega_{2}}-1\right\} e^{\alpha\left|z_{1}-z_{2}\right|}-\cos\left(\Phi_{0}\right)}\right| \leq \frac{1+\left| \frac{\alpha\left(U_{1}-U_{2}\right)}{\upomega_{1}+\upomega_{2}}-1\right| e^{\alpha\left|z_{1}-z_{2}\right|}}{\left|\left| \frac{\alpha\left(U_{1}-U_{2}\right)}{\upomega_{1}+\upomega_{2}}-1\right| e^{\alpha\left|z_{1}-z_{2}\right|}-1\right|}=G_{opt}^{global},
\end{equation}
The optimal gain is maximum when $\Phi_{0}=0$, and is given by 
\begin{equation}
%G_{opt}=\left|\frac{\left\{ \frac{\alpha\left(U_{1}-U_{2}\right)}{\upomega_{1}+\upomega_{2}}-1\right\} e^{\alpha\left|z_{1}-z_{2}\right|}+\cos\left(\Phi_{0}\right)}{\left\{ \frac{\alpha\left(U_{1}-U_{2}\right)}{\upomega_{1}+\upomega_{2}}-1\right\} e^{\alpha\left|z_{1}-z_{2}\right|}-\cos\left(\Phi_{0}\right)}\right| \leq \frac{1+\left| \frac{\alpha\left(U_{1}-U_{2}\right)}{\upomega_{1}+\upomega_{2}}-1\right| e^{\alpha\left|z_{1}-z_{2}\right|}}{\left|\left| \frac{\alpha\left(U_{1}-U_{2}\right)}{\upomega_{1}+\upomega_{2}}-1\right| e^{\alpha\left|z_{1}-z_{2}\right|}-1\right|}=G_{opt}^{global},
G_{opt}^{global}=\frac{\left|Q\right|+1}{\left|Q\right|-1}.
\end{equation}
 $G_{opt}^{global}$ is known as the \emph{global optimal gain}. We take KH instability as an example for studying the  non-modal behaviour. The condition $|Q| > 1$ in this case translates to  $\alpha > 0.64$ (see  \S \ref{sec:KH}).  From  figure \ref{fig:manyfigs}(c) we find that the perturbation amplitude   corresponding to  $\alpha=0.65$ grows  to a maximum of  $\sqrt{G_{opt}^{global}}=4.56$ times its initial value. Similar results have been obtained by \cite{heif2005}, who studied the non-normal behaviour using the enstrophy norm.  Figure \ref{fig:manyfigs}(d) shows the corresponding temporal variation of its phase-shift.  Unlike figure \ref{fig:Summaryfig}(a), there is no phase-locking since N\&S condition is not satisfied here. Instead, the two ``marginally stable'' (as referred to in the classical theory) wave-modes continue to propagate in opposite directions. Hence the phase-shift oscillates between $0$ and $2\pi$. 
 
\section{Analysis of classical shear instabilities using WIT}
\label{sec:inst}

The  WIT formulation proposed in \S \ref{sec:waveint} is quite general in the sense that we have not prescribed the types of the constituent waves. Different kinds of shear instabilities may arise depending on the wave types. For example, the interaction between two vorticity waves results in KH instability; Taylor-Caulfield instability results from the  interaction
between two gravity waves in constant shear, while the interaction between a vorticity and a gravity wave produces Holmboe instability.   In this section we will use WIT to analyze  the three above mentioned  classical examples of shear instabilities. 

\subsection{The Rayleigh/Kelvin-Helmholtz Instability}
\label{sec:KH}

\begin{figure}

\centering
\subfloat[]{\includegraphics[scale=0.064]{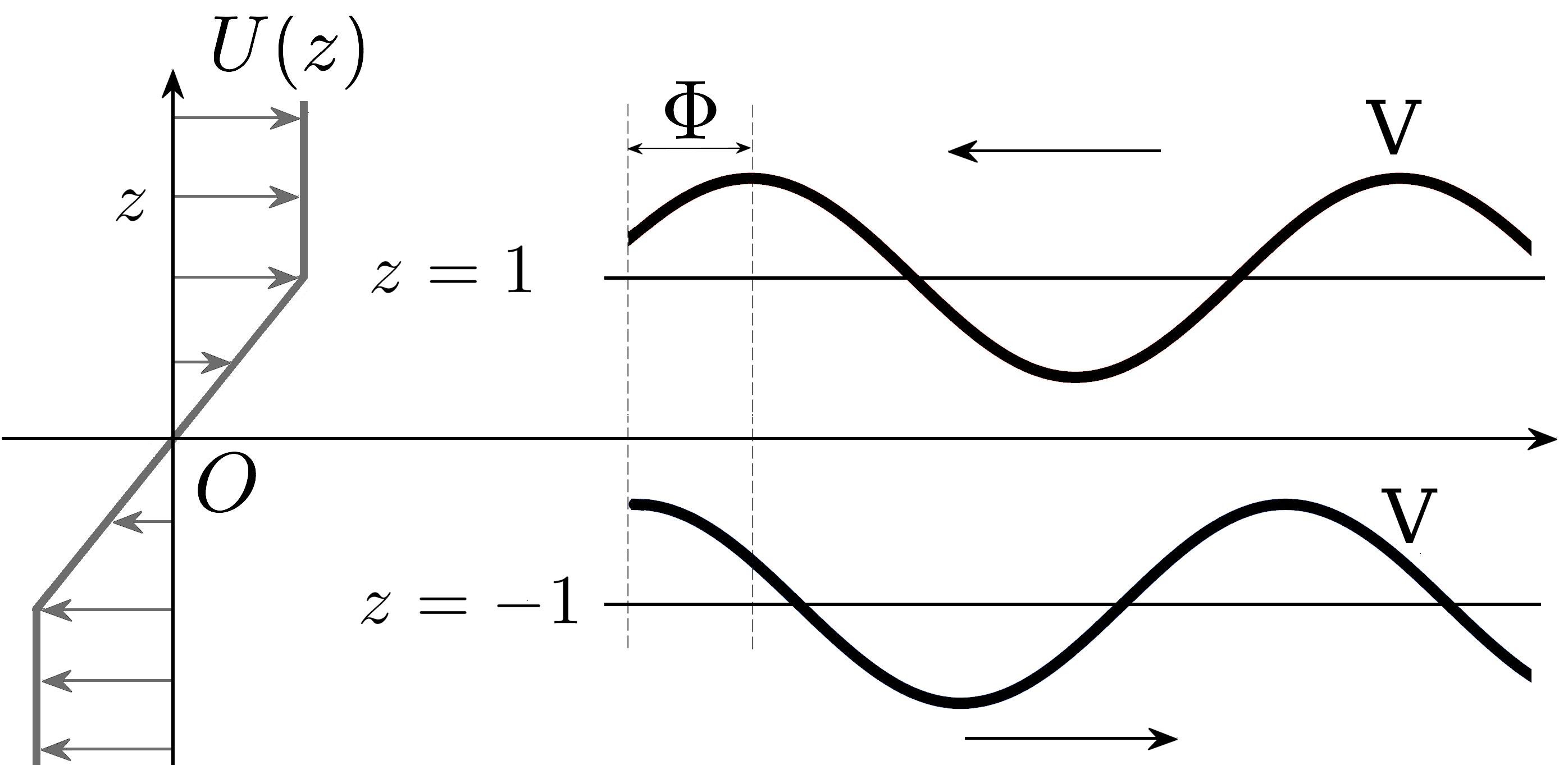}
}
\centering
\subfloat[]{\includegraphics[scale=0.043]{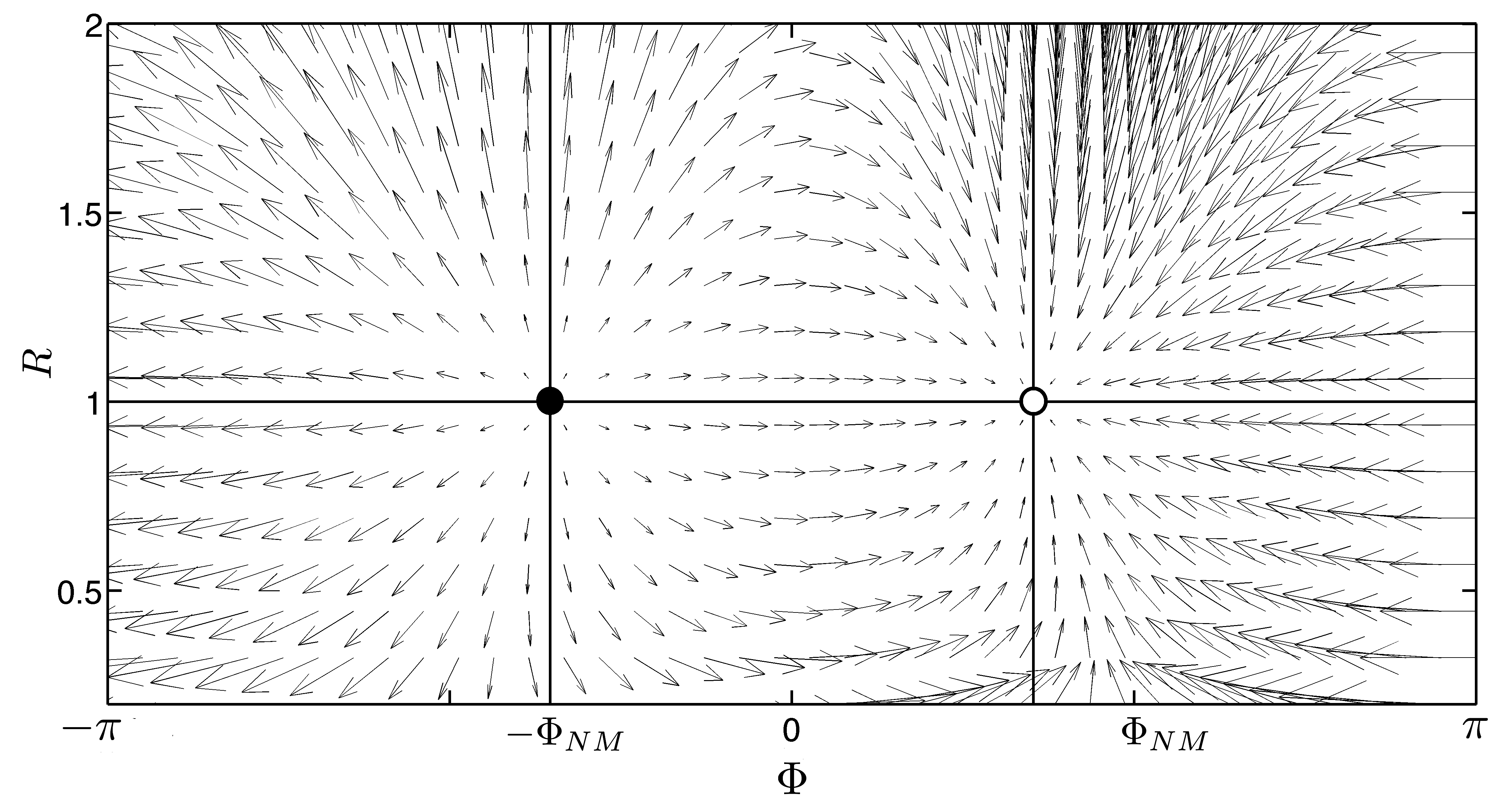}
}
\setlength{\belowcaptionskip}{-20pt}
\caption{(a) The setting leading to the KH instability. The velocity profile in  (\ref{eq:kh2}) is shown on the left, while the vorticity waves (marked by ``V'') are shown on the right. (b) Phase portrait of KH instability corresponding to $\alpha=0.4$. The system has two equilibrium points - the source node (\textbullet) and the sink node $(\circ)$.  \bigskip{}}
\label{fig:KH}
\end{figure}

Let us consider a piecewise linear velocity profile
 \begin{equation}  
U\left(z\right)=\begin{cases}
U_{1} & \,\,\,\,\,\,\,\,\,\,\,\,\,\,z\geq z_{1}\\
Sz & z_{2}\leq  z \leq z_{1}\\
U_{2} &  \,\,\,\,\,\,\,\,\,\,\,\,\,z \leq z_{2}. \end{cases}\label{eq:KH}\end{equation}
This  profile is a prototype of barotropic shear layers occurring in many geophysical and astrophysical flows \cite[]{guha2012}. It supports two vorticity waves, one  at $z_{1}$ and the other at $z_{2}$. The shear $S=(U_{1} -U_{2})/ (z_{1}-z_{2})$. We nondimensionalize 
the problem by choosing a  length scale $h=(z_{1} -z_{2})/2$ and a velocity scale
 $\Delta U=(U_{1} -U_{2})/2$. In a reference frame moving with the mean flow $\bar{U}=(U_{1}+U_{2})/2$,   the non-dimensional velocity profile becomes
 \begin{equation}
U\left(z\right)=\begin{cases}
\,\,\,\,1 & \,\,\,\,\,\,\,\,\,\,\,\,\,\,\,\,z\geq1\\
\,\,\,\,z & -1\leq z\leq1\\
-1 & \,\,\,\,\,\,\,\,\,\,\,\,\,\,\,z\leq-1.\end{cases}\label{eq:kh2}\end{equation}
This profile, along with the vorticity waves, is shown in figure\ \ref{fig:KH}(a). The top wave is left moving while the bottom wave moves to the right.
%Both the waves counter-propagate, i.e.\ move in a direction opposite to the background flow. 
% (), and therefore their phase-speed gets reduced.  Further reduction in  phase-speed occurs through wave interaction - the vertical velocity 
% field produced by the distant wave acts so as to diminish the phase-speed of the given wave. This continues until the two waves get   ``phase-locked'', i.e.\ they are stationary relative to each other. In this phase-locked state,
% the two waves grow exponentially fast (i.e.\  pure modal growth) until the shear layer becomes non-linear, producing elliptical vortices \cite[]{guha2012}. 

Using the classical normal-mode theory \cite{rayl1880} showed that the profile in  (\ref{eq:kh2}) is linearly unstable for  $0 \leq \alpha \leq 0.64$. This instability is often referred to as the ``Rayleigh's shear instability''. However we have addressed it as the ``Rayleigh/Kelvin-Helmholtz'' instability, and used the acronym ``KH''.  

Figure \ref{fig:KH}(a) reveals that the  KH instability can be analyzed using  WIT framework.  Since the two waves involved in the KH instability are vorticity waves, we substitute  (\ref{eq:vortwaverel}) in  (\ref{eq:27})-(\ref{eq:30}), and after non-dimensionalization we obtain
\begin{eqnarray}
 &  & \gamma_{1}=\frac{1}{2R} e^{-2\alpha}\sin\Phi \label{eq:khh1}\\
 &  & c_{1}=1-\frac{1}{2\alpha}\left[1-\frac{1}{R}e^{-2\alpha}\cos\Phi\right]\label{eq:khh2}\\
 &  & \gamma_{2}=\frac{R}{2}e^{-2\alpha}\sin\Phi\label{eq:khh3}\\
 &  & c_{2}=-1+\frac{1}{2\alpha}\left[1-Re^{-2\alpha}\cos\Phi\right].\label{eq:khh4}
 \end{eqnarray}
Eqs.\ (\ref{eq:khh1})-(\ref{eq:khh4}) provide a \emph{non-modal} description of KH instability. The equation-set is isomorphic to (14a)-(14d) of \cite{heif1999} and homomorphic to  (7a)-(7d) of \cite{davies1994}.
 While the equation-set described by \cite{heif1999} shows how counter-propagating Rossby wave 
interactions lead to barotropic shear instability,  the equation-set  formulated by \cite{davies1994} shows how baroclinic instability is produced through the interaction of temperature edge waves.
%of the Eady model. Furthermore, \cite{heif1999}  showed that
%their set of equations is homomorphic to that of \cite{davies1994}.
%
% The fact that wave interaction modifies the phase-speed of a vorticity wave
%can be understood by comparing  (\ref{eq:khh2}) and  (\ref{eq:khh4}) with the non-dimensional form of  (\ref{eq:vortspeed}) (nondimensionalization means substituting $S=1$
%and $U_{i}=1$ or $-1$ in  (\ref{eq:vortspeed})).  

The generalized non-linear dynamical system given by (\ref{eq:dRdt})-(\ref{eq:dPhidt}) in this case translates to 
\begin{eqnarray}
& & \dot{R}=\frac{1}{2}\left(1-R^{2}\right)e^{-2\alpha} \sin\Phi \label{eq:kh_dyn1}\\
& & \dot{\Phi}=\left(2\alpha-1\right)+\frac{1}{2}\left(R+\frac{1}{R}\right)e^{-2\alpha}\cos\Phi.  \label{eq:kh_dyn2}
\end{eqnarray}
The equilibrium points  of this system are $(R_{NM},\pm \Phi_{NM})$, where
\begin{eqnarray}
 &  & R_{NM}=1\\
 &  & \Phi_{NM}=\cos^{-1}\left[\left(1-2\alpha\right)e^{2\alpha}\right].
 \label{eq:nor_mode_KH}
\end{eqnarray}

The phase portrait is shown in  figure\ \ref{fig:KH}(b). It confirms that the dynamical system is indeed of source-sink type, as predicted in \S \ref{sec:waveint}. 

The N\&S condition for instability expressed via  (\ref{eq:necessary}) in this case reads
\begin{equation}
-1\leq\left(1-2\alpha\right)e^{2\alpha}\leq1\,\,\,\,\mathsf{\mathrm{implying}}\,\,\,\,0 \leq \alpha \leq 0.64.
\end{equation}
The range of unstable wavenumbers obtained  from the above equation corroborates Rayleigh's normal-mode analysis. Normal-mode theory also shows that the
 wavenumber of maximum growth is $\alpha_{max}=0.4$.  The same answer can be obtained from WIT by imposing the normal-mode condition and maximizing $ \gamma_{1}$ or  $\gamma_{2}$ with respect to  $\alpha$.

The fact that KH instability develops into a standing wave instability can be verified by applying the normal-mode condition in   (\ref{eq:khh2}) and (\ref{eq:khh4}). Performing the necessary steps we find $c_{1}=c_{2}=0$, i.e.\ the waves  became stationary after phase-locking. The waves stay in this configuration and grow exponentially, hence the shear layer grows in size. The growth process eventually becomes non-linear, and the  shear layer modifies into elliptical patches of constant vorticity  \cite[]{guha2012}.

%\begin{figure}
%\centering
%\includegraphics[scale=0.08]{KH_phase}
%\setlength{\belowcaptionskip}{-20pt}
% \caption{Phase portrait of Rayleigh/Kelvin-Helmholtz instability corresponding to $\alpha=0.4$. The system has two equilibrium points - one unstable $(\circ)$ and the other stable (\textbullet). $\Phi$ is the phase difference between the 
% lower and upper waves, while $R$ represents the ratio of the upper wave amplitude to the lower wave amplitude.\bigskip{}}
%\label{fig:KHphase}
%\end{figure}

%%%%%%%%%%%%%%%%%%%%%%%

\subsection{The Taylor-Caulfield Instability}

\begin{figure}

\centering
\subfloat[]{\includegraphics[scale=0.06]{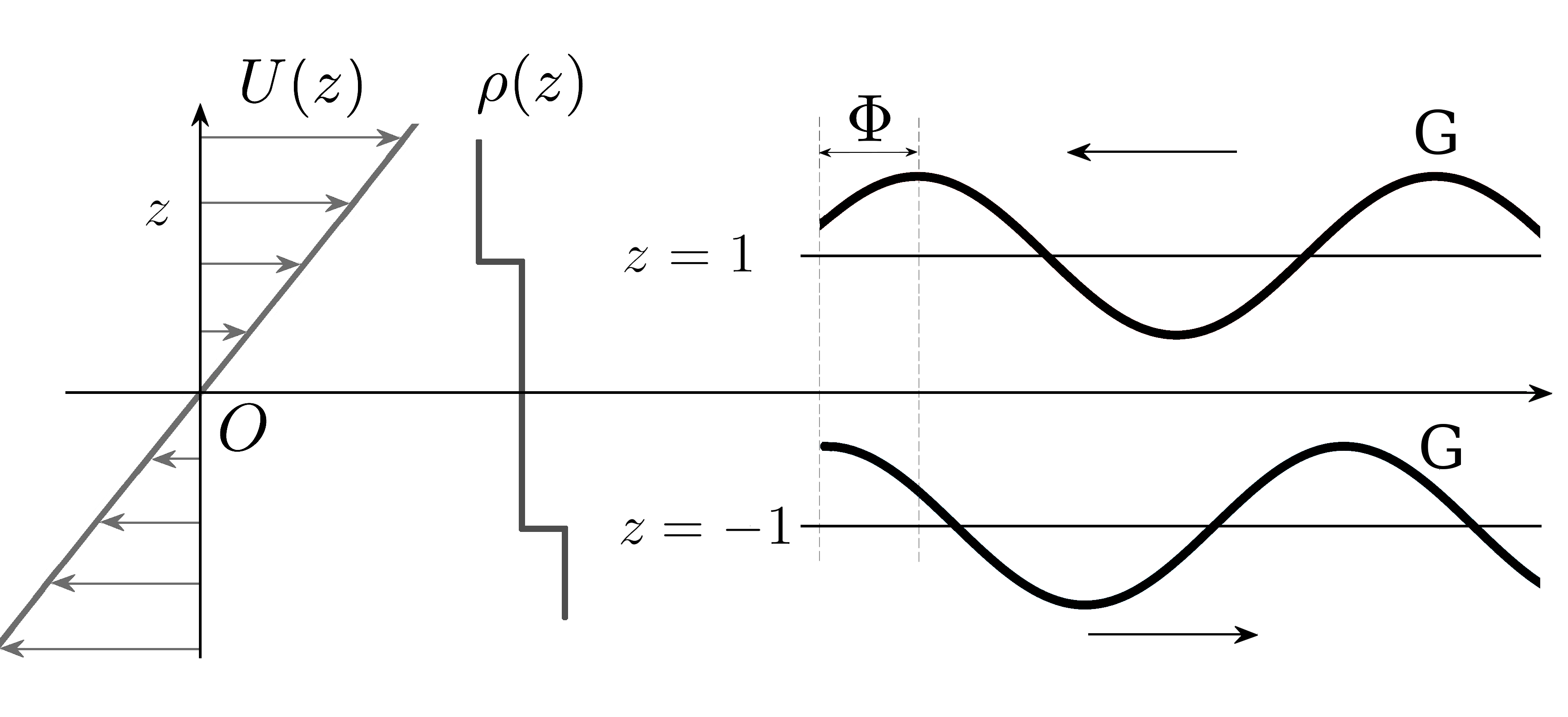}

}

\centering
\subfloat[]{\includegraphics[scale=0.042]{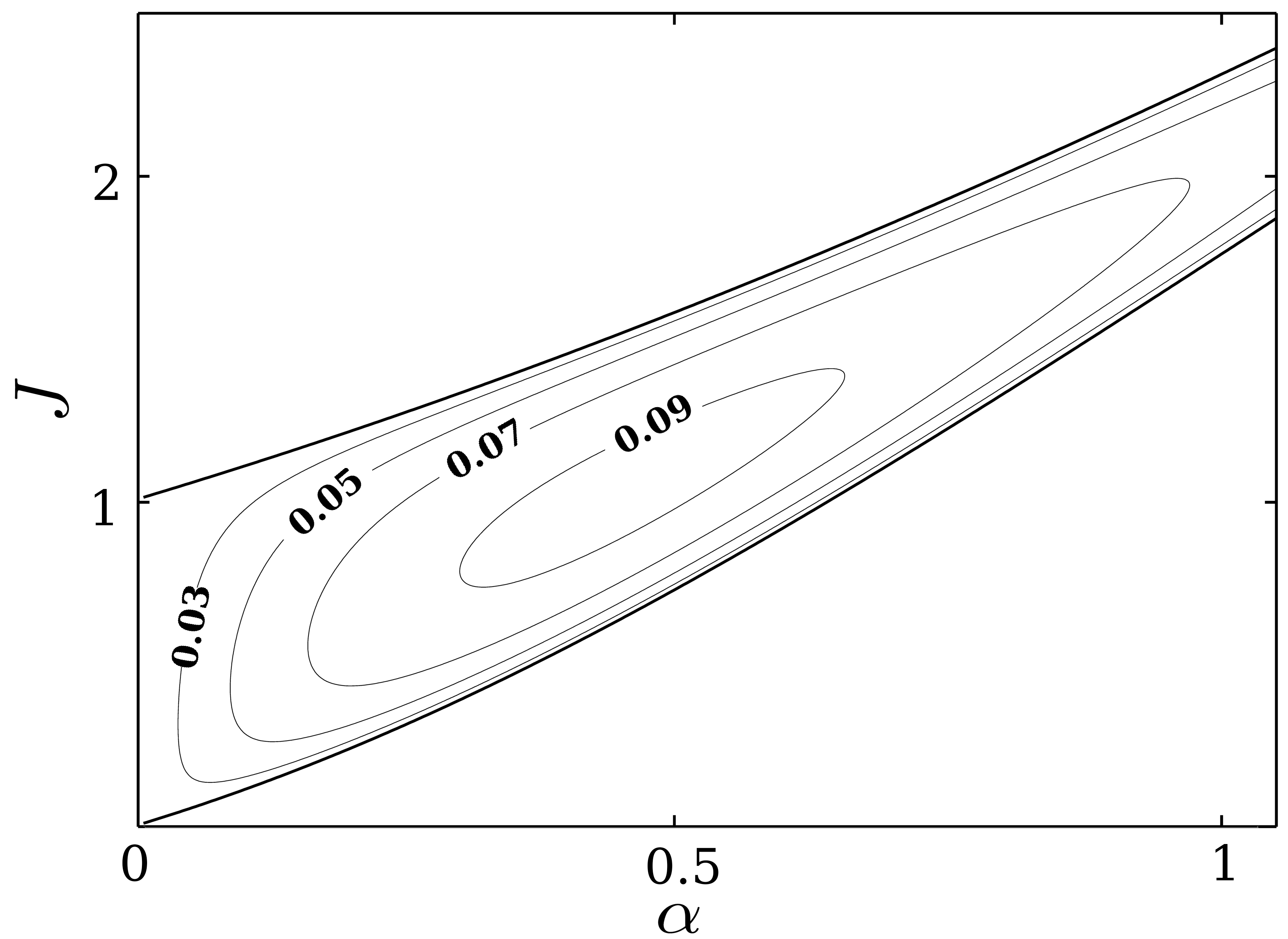}

}
\centering
\subfloat[]{\includegraphics[scale=0.047]{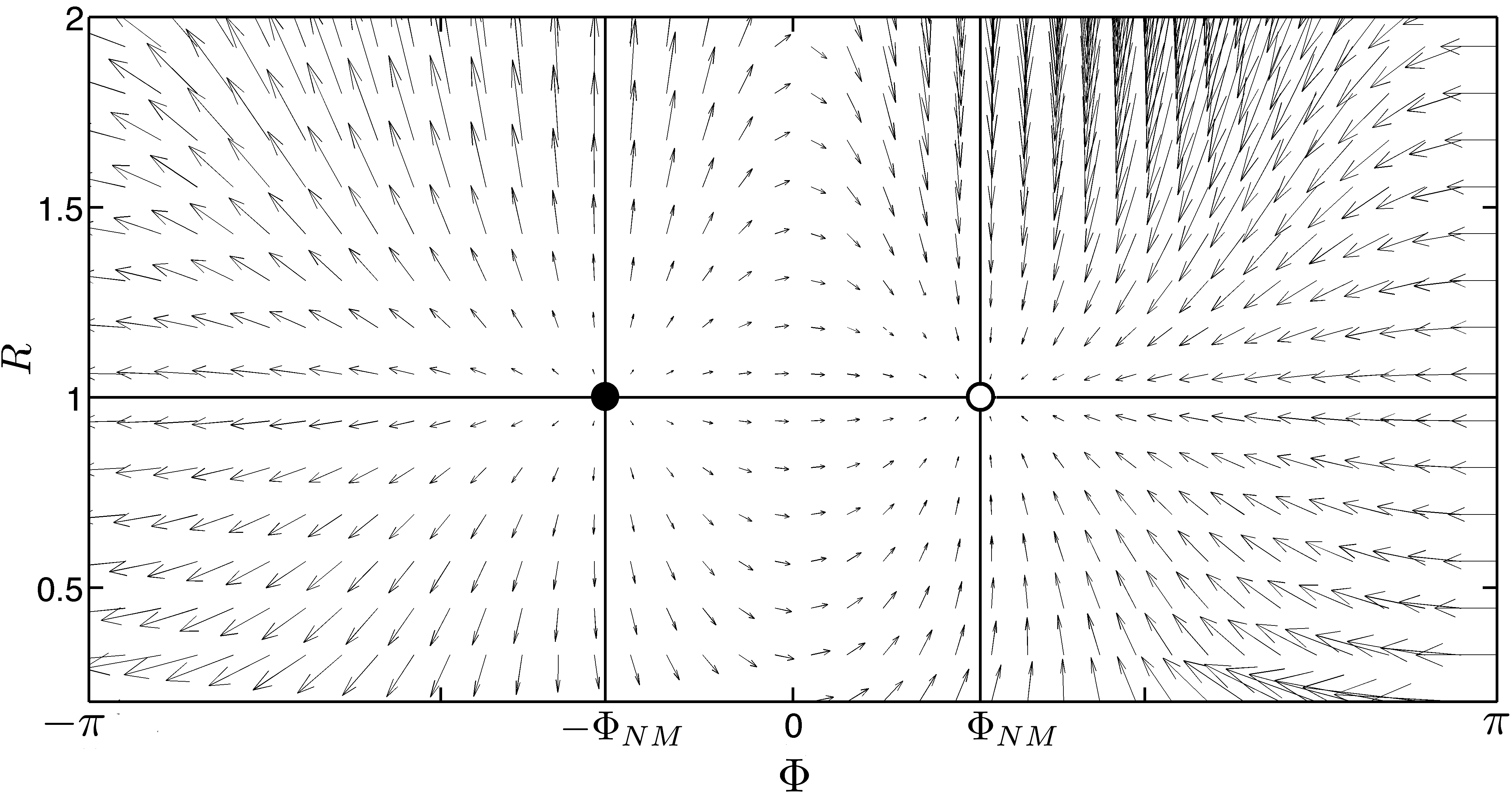}

}
\setlength{\belowcaptionskip}{-50pt}
\caption{(a) The setting leading to the TC instability. The velocity and density profiles in  (\ref{eq:tayl1}) are shown on the left, while the gravity waves ``G''  are shown on the right. (b) Linear stability diagram of the 
TC instability. The contours represent the growth-rate. (c) Phase portrait of TC instability corresponding to an unstable combination of $\alpha$ and $J$ ($\alpha=0.2$ and $J=0.7264$)\bigskip{}}
\label{fig:Taylor}
\end{figure}

 Let us consider a uniform shear layer with two density interfaces
 \begin{equation}
U(z)=Sz\,\,\,\,\,\,\,\,\mathrm{and}\,\,\,\,\,\,\,\,\rho\left(z\right)=\begin{cases}
\rho_{0} -\frac{\Delta \rho}{2} & \,\,\,\,\,\,\,\,\,\,\,\,\,\,z\geq z_{1}\\
\rho_{0} &  z_{2}\leq z\leq z_{1}\\
\rho_{0} + \frac{\Delta \rho}{2} & \,\,\,\,\,\,\,\,\,\,\,\,\,z\leq z_{2}.
\end{cases}
\label{eq:tayl0}
\end{equation}
The shear $S$ is constant. We choose $\Delta \rho/2$ as the density scale, $h=(z_{1} -z_{2})/2$ as the length scale, and thereby nondimensionalize  (\ref{eq:tayl0}). 
The physical state of the system is determined by the competition between  density stratification and shear, the non-dimensional measure of which is given by the Bulk 
Richardson number $J = g'/(h S^{2})$, where  $g'=g(\Delta \rho/ 2)/\rho_{0}$ is the reduced gravity, and $\rho_{0}$ is the reference density.  The dimensionless velocity and density profiles therefore become 
\begin{equation}
U(z)=z\,\,\,\,\,\,\,\,\mathrm{and}\,\,\,\,\,\,\,\,\,\rho\left(z\right)=\begin{cases}
-1 & \,\,\,\,\,\,\,\,\,\,\,\,\,\,\,\,z\geq1\\
\,\,\,\,0 & -1\leq z\leq1\\
\,\,\,\,1 & \,\,\,\,\,\,\,\,\,\,\,\,\,\,\,\,z\leq-1.
\end{cases}
\label{eq:tayl1}
\end{equation}
This flow configuration is shown in figure\ \ref{fig:Taylor}(a). Notice that a \emph{homogeneous} flow with $U(z)=z$ is linearly stable. Moreover, a \emph{stationary} fluid having a density distribution as in (\ref{eq:tayl1})  is gravitationally stable. Surprisingly, the flow obtained by superimposing the two stable configurations is linearly unstable. \cite{tayl1931}  was the first to report and provide a detailed theoretical analysis of this rather non-intuitive instability.  The first experimental evidence of this instability was provided by \cite{caul1995}. Following \cite{carp2012} we refer to it as the ``Taylor-Caulfield (TC) instability''.  The interplay between the background shear and the gravity waves existing at the density interfaces produce the destabilizing effect. 
 \cite{tayl1931} found that for each value of $J$, there exists a band of unstable wavenumbers (and vice-versa), shown in figure\ \ref{fig:Taylor}(b). 
This unstable range is given by (see \cite{hm73} or  (2.154) of  \cite{suth2010}):
\begin{equation}
\frac{2\alpha}{1+e^{-2\alpha}}\leq J\leq\frac{2\alpha}{1-e^{-2\alpha}}.
\label{eq:ta1}
\end{equation}

Probably the only way to make sense of TC instability is through wave interactions, which has been described in 
 \cite{caul1994} and \cite{carp2012} using normal-mode theory. %A brief explanation of TC is as follows: From \S \ref{subsec:gravitywaves} we found that each density interface  (located at $z=1$ and $z=-1$) supports two gravity waves. The background flow is such that the left moving gravity wave at the upper interface is Doppler-shifted towards right, while the right moving gravity wave at the lower interface is Doppler-shifted towards left.  
 WIT provides  the framework for studying non-modal TC instability. Substituting  (\ref{eq:gravwaverel}) in  (\ref{eq:27})-(\ref{eq:30}), and performing non-dimensionalization, we obtain  
\begin{eqnarray}
 &  & \gamma_{1}=\frac{J}{2R(1+c_{2})}e^{-2\alpha}\sin\Phi \label{eq:T1}\\
 &  & c_{1}=1-\sqrt{\frac{J}{2\alpha}\left(1-\frac{\beta}{R}e^{-2\alpha}\cos\Phi\right)} \label{eq:T2}\\
 &  & \gamma_{2}=\frac{JR}{2(1-c_{1})}e^{-2\alpha}\sin\Phi \label{eq:T3}\\
 &  & c_{2}=-1+\sqrt{\frac{J}{2\alpha}\left(1-\frac{R}{\beta}e^{-2\alpha}\cos\Phi\right)}. \label{eq:T4}
\end{eqnarray}
Here  $\beta=\upomega_{2}/\upomega_{1}=(1-c_{1})/(1+c_{2})$, and by definition is a positive quantity. 
 From (\ref{eq:T2}) and (\ref{eq:T4}) we construct a quadratic equation for $\beta$:
\begin{equation}
\beta^{2}+\beta e^{-2\alpha}\cos\Phi\left(\frac{1}{R}-R\right)-1=0.
\label{eq:betaevol}
\end{equation}
Amongst the two roots, only the positive root is relevant.  Equations (\ref{eq:T1})-(\ref{eq:T4}) provide a \emph{non-modal} description of TC instability. The equation-set is of coupled nature, and is therefore \emph{not} homomorphic to (\ref{eq:27})-(\ref{eq:30}).  The non-linear dynamical system in this case is given by
\begin{eqnarray}
 &  & \dot{R}=\frac{J}{2}\left(\frac{1}{1+c_{2}}-\frac{R^{2}}{1-c_{1}}\right)e^{-2\alpha}\sin\Phi \label{eq:TR}\\
 &  & \dot{\Phi}=2\alpha-\sqrt{\frac{J\alpha}{2}\left(1-\frac{\beta}{R}e^{-2\alpha}\cos\Phi\right)}-\sqrt{\frac{J\alpha}{2}\left(1-\frac{R}{\beta}e^{-2\alpha}\cos\Phi\right)}. \label{eq:Tphi}
\end{eqnarray}

%\begin{figure}
%\centering
%\includegraphics[scale=0.07]{Taylor_phase}
%\caption{Phase portrait of Taylor-Caulfield instability corresponding to an unstable combination of $\alpha$ and $J$. Here $\alpha=0.2$ and $J=0.7264$.}
%\label{fig:Taylphase}
%\end{figure}

%It is plain to recognize the qualitative similarity between this phase portrait and that of KH. This similarity is due to the fact that the wave
%interaction process for Taylor instability works in a manner  analogous to KH, the only difference being that the two waves here are gravity waves. 
At phase-locking $R=R_{NM}=\sqrt{\beta}$. Substituting this value in  (\ref{eq:betaevol}) gives $\beta=1$. Therefore $R_{NM}=1$ and $c_{1}=c_{2}=0$ at resonance, which implies that TC instability, like KH, also evolves into a
stationary disturbances. 

The phase portrait is shown in figure \ref{fig:Taylor}(c). The phase-shift $\Phi_{NM}$ is evaluated from  (\ref{eq:Tphi}):
\begin{equation}
\Phi_{NM}=\cos^{-1}\left[\left(1-\frac{2\alpha}{J}\right)e^{2\alpha}\right].
\label{eq:nor_mode_taylor}
\end{equation}

The N\&S condition for TC instability is given by 
\begin{equation}
-1\leq\left(1-\frac{2\alpha}{J}\right)e^{2\alpha}\leq1\,\,\,\,\mathsf{\mathrm{implying}}\,\,\,\, \frac{2\alpha}{1+e^{-2\alpha}}\leq J\leq\frac{2\alpha}{1-e^{-2\alpha}}.
\end{equation}
The latter result corroborates the classical normal-mode result given in  (\ref{eq:ta1}). 

Non-modal TC instability has also been studied  by \cite{rabi2011}.  In their paper the non-modal equation-set is given by (3.11a,b)-(3.12a,b), which is quite different from our equation-set (\ref{eq:T1})-(\ref{eq:T4}). However, these two equation-sets should be identical since they are describing the same physical problem. Using the equation-set of \cite{rabi2011} we obtained the following range of unstable wavenumbers: $2\alpha/\left(1+e^{-2\alpha}\right)^{2} \leq J\leq2\alpha/\left(1-e^{-2\alpha}\right)^{2}$. This bound is different from the correct stability bound (\ref{eq:ta1}). %it is not possible to verify the stability boundary (\ref{eq:ta1}) by using the equation-set of \cite{rabi2011}. 
%the equation-set (3.11a,b)-(3.12a,b) of  \cite{rabi2011} is erroneous since it does not produce the correct stability boundary given by (\ref{eq:ta1}). The range of unstable wavenumbers obtained from their equation-set is . Notice the appearance of an extra term 

\subsection{The Holmboe Instability}

\begin{figure}

\centering
\subfloat[]{\includegraphics[scale=0.06]{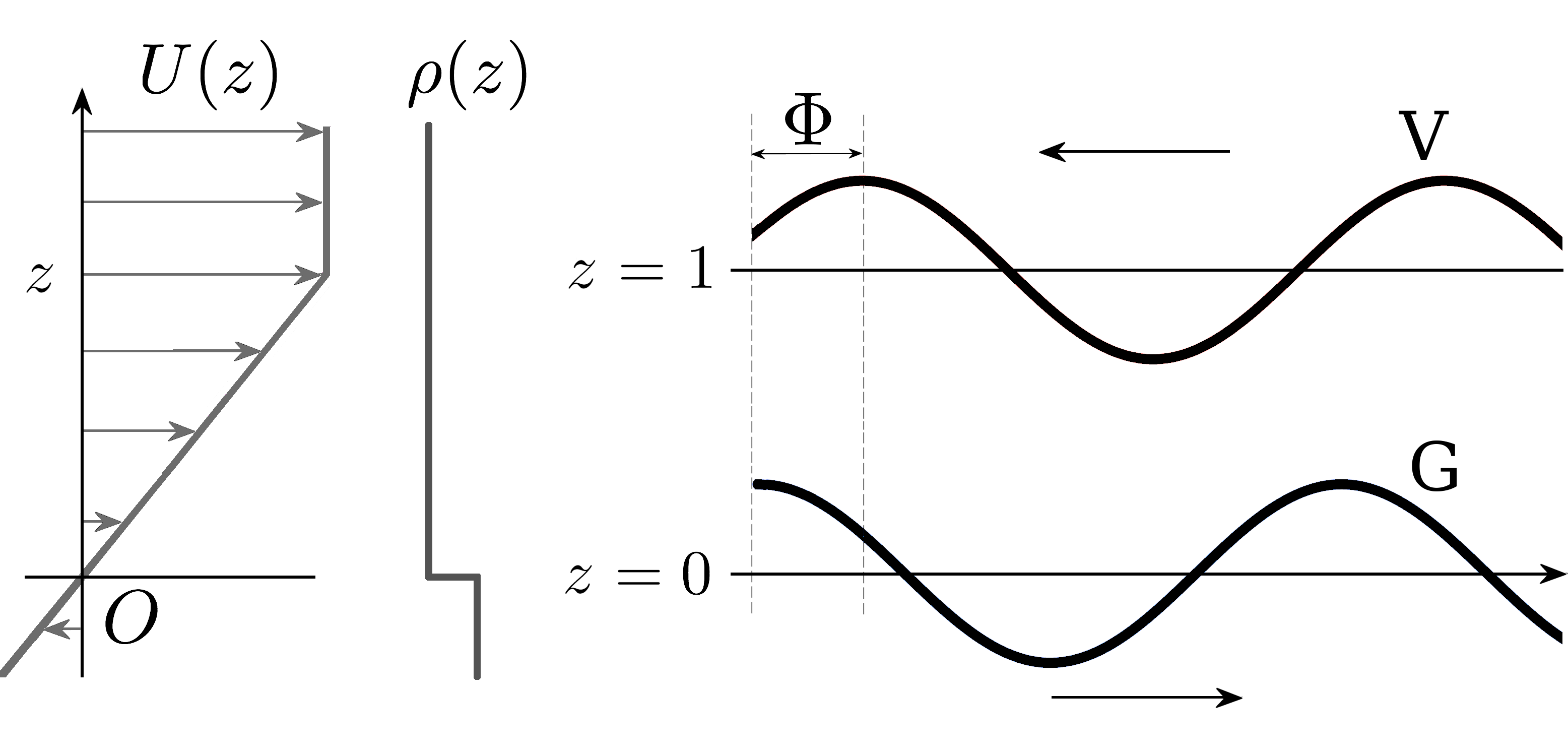}
}

\centering
\subfloat[]{\includegraphics[scale=0.042]{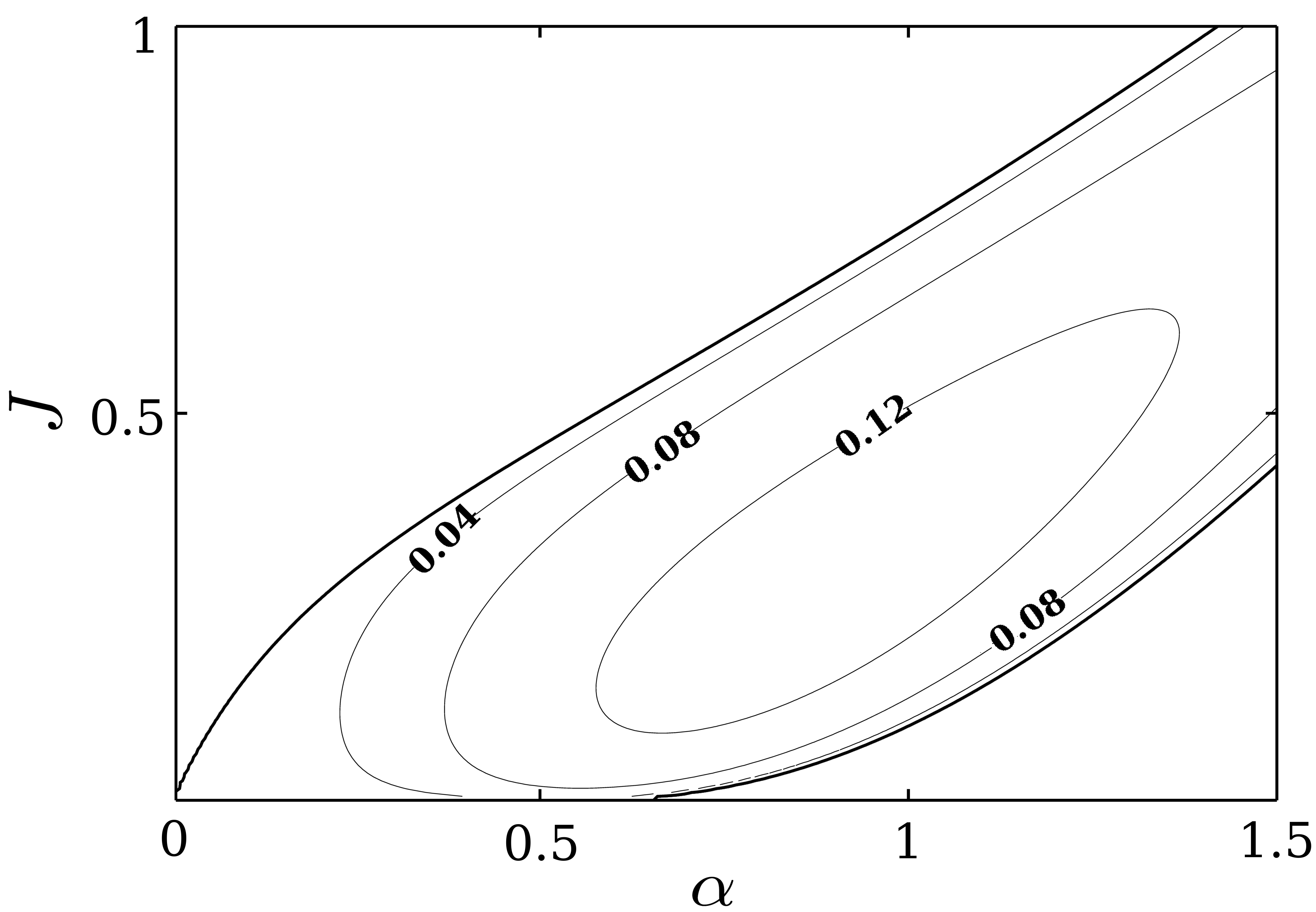}
}
\centering
\subfloat[]{\includegraphics[scale=0.045]{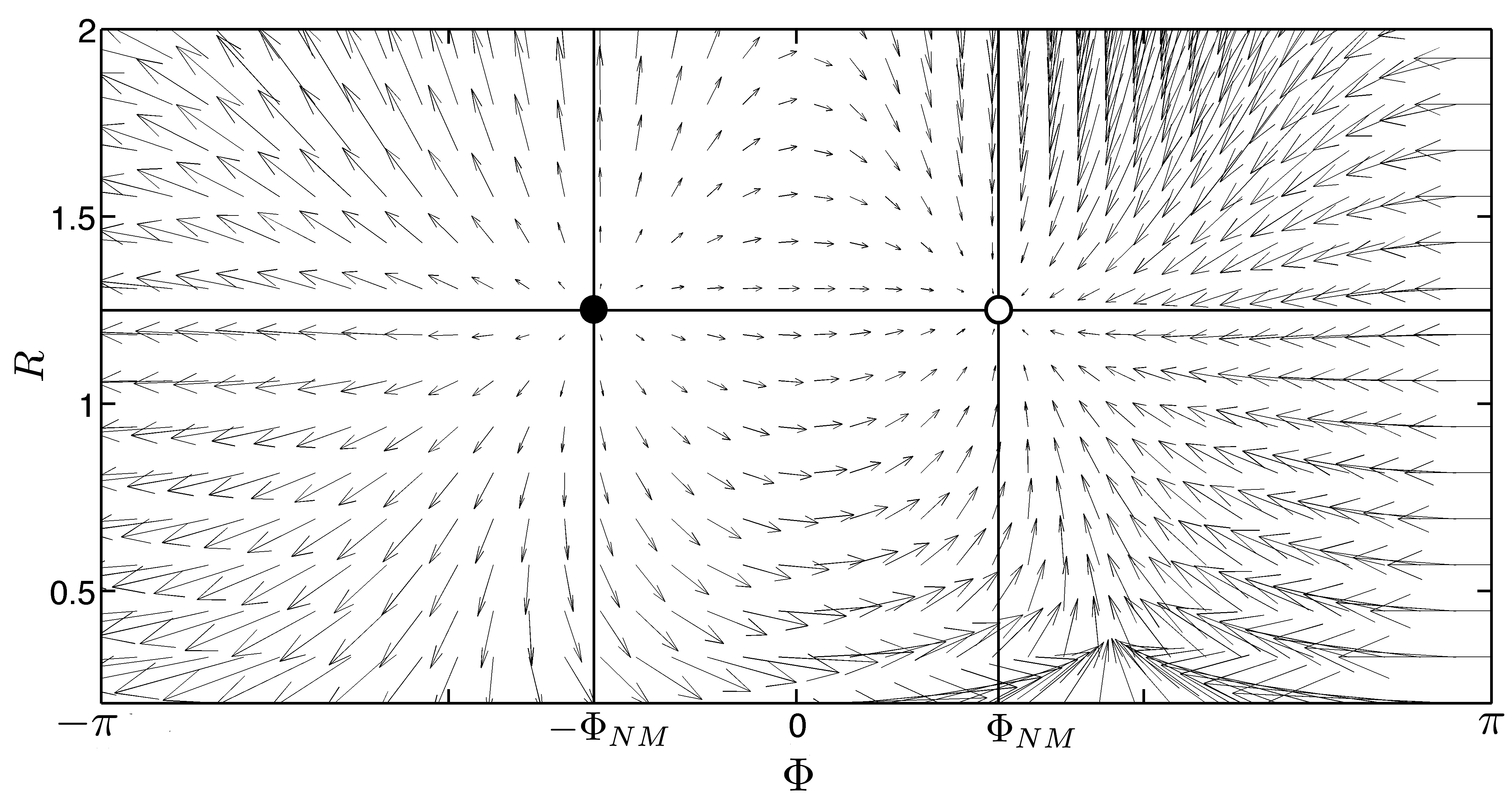}
}
\setlength{\belowcaptionskip}{-20pt}
\caption{(a) The setting leading to the Holmboe instability. The velocity and density profiles in  (\ref{eq:holm1}) are shown on the left, while the vorticity wave ``V'' and the gravity wave ``G''  are shown on the right. (b) 
Linear stability diagram; contours represent  growth-rate. (c) Phase portrait corresponding to $\alpha=1$ and $J=0.5$.\bigskip{}}

\label{fig:Holmboe}

\end{figure}

Let us consider the following velocity and density profiles
\begin{equation}
U(z)=\begin{cases}
U_{1} & z\geq z_{1}\\
Sz & z\leq z_{1}
\end{cases}\,\,\,\, \mathrm{and}\,\,\,\,\rho\left(z\right)=\begin{cases}
\rho_{0} & z\geq z_{2}\\
\rho_{0}+\triangle\rho & z\leq z_{2}.
\end{cases}
\label{eq:holm0}
\end{equation}
We nondimensionalize  (\ref{eq:holm0}) exactly like the TC instability problem, which gives us  the dimensionless velocity and density profiles:
\begin{equation}
U(z)=\begin{cases}
1 & z\geq1\\
z & z\leq1
\end{cases}\,\,\,\, \mathrm{and}\,\,\,\,\rho\left(z\right)=\begin{cases}
0 & z\geq0\\
2 & z\leq0.
\end{cases}
\label{eq:holm1}
\end{equation}
The vorticity interface at the top supports a vorticity wave, while the density interface at the bottom supports two gravity waves. 
The interaction between the left moving vorticity wave at the upper interface and the right moving gravity wave at the lower interface leads to an
instability mechanism, known as the ``Holmboe instability'' (Note that the generation of ocean surface gravity waves by wind shear \cite[]{miles1957} can be viewed as  ``non-Boussinesq Holmboe instability'', and can therefore be interpreted using wave interactions.). The corresponding flow setting is shown in figure\ \ref{fig:Holmboe}(a). 

\cite{holm1962}  was the first to consider the instability mechanism resulting from the interaction between vorticity and gravity waves. Performing a normal-mode stability analysis, Holmboe discovered the existence of an unstable mode characterized by traveling waves. The profile used by Holmboe is more complicated than that which we are considering. Holmboe's profile, which included multiple wave interactions, was substantially simplified by \cite{bain1994} by introducing the profile in (\ref{eq:holm1}).  Linear stability analysis shows that corresponding to each value of $J$, there exists a band of unstable wavenumbers, shown in figure\ \ref{fig:Holmboe}(b). The stability boundary has been evaluated in   Appendix \ref{app:A}.

In order to understand the Holmboe instability in terms of WIT,  we  substitute  (\ref{eq:vortwaverel}) and   (\ref{eq:gravwaverel})  in  (\ref{eq:27})-(\ref{eq:30}). 
After   non-dimensionalization, we obtain  
\begin{eqnarray}
 &  & \gamma_{1}=\frac{J}{Rc_{2}}e^{-\alpha}\sin\Phi \label{eq:H1}\\
 &  & c_{1}=1-\frac{1}{\alpha}\left(\frac{1}{2}-\frac{J}{Rc_{2}}e^{-\alpha}\cos\Phi\right) \label{eq:H2}\\
 &  & \gamma_{2}=\frac{R}{2}e^{-\alpha}\sin\Phi \label{eq:H3}\\
 &  & c_{2}=\frac{1}{4\alpha}\left(-Re^{-\alpha}\cos\Phi+\sqrt{R^{2}e^{-2\alpha}\cos^{2}\Phi+16\alpha J}\right). \label{eq:H4}
\end{eqnarray}
Like the TC instability case, the Holmboe equation-set is also of coupled nature, and is therefore not homomorphic to (\ref{eq:27})-(\ref{eq:30}). The  non-linear dynamical system is given by:
\begin{eqnarray}
 &  & \dot{R}=\left(\frac{4\alpha J}{-Re^{-\alpha}\cos\Phi+\sqrt{R^{2}e^{-2\alpha}\cos^{2}\Phi+16\alpha J}}-\frac{R^{2}}{2}\right)e^{-\alpha}\sin\Phi \label{eq:HR} \\
 &  & \dot{\Phi}=\alpha-\frac{1}{2}\left(1-Re^{-\alpha}\cos\Phi\right)+ \nonumber \\
 &  & \,\,\,\,\,\,\,\,\,\,\,\,\,\,\,\,\frac{4\alpha J}{-Re^{-\alpha}\cos\Phi+\sqrt{R^{2}e^{-2\alpha}\cos^{2}\Phi+16\alpha J}}\left(\frac{e^{-\alpha}\cos\Phi}{R}-1\right). \label{eq:HPhi}
\end{eqnarray}
The  equilibrium points of this system are $(R_{NM},\pm \Phi_{NM})$, where
\begin{eqnarray}
&  & R_{NM}=\sqrt{\frac{1-2\alpha+\sqrt{32\alpha J+\left(1-2\alpha\right)^{2}}}{2}} \label{eq:Rnm}\\
&  &  \Phi_{NM}=\cos^{-1}\left[\left(\frac{R_{NM}^{2}+1-2\alpha}{2R_{NM}}\right)e^{\alpha}\right].  \label{eq:nor_mode_holmboe}
\end{eqnarray}

The  N\&S condition for Holmboe instability is found to be
\begin{equation}
-1\leq\left(\frac{R_{NM}^{2}+1-2\alpha}{2R_{NM}}\right)e^{\alpha}\leq1. 
\end{equation}
This provides the range of $J$ leading to Holmboe instability, and is as follows:
\begin{equation}
\frac{1}{2A}\left(-B-\sqrt{B^{2}-4AC}\right)\leq J\leq\frac{1}{2A}\left(-B+\sqrt{B^{2}-4AC}\right),
\label{eq:bound_holmboe}
\end{equation}
where
\begin{eqnarray*}
 &  & A=16\alpha^{2}\\
 &  & B=-\alpha\left[8\left(2\alpha-1\right)^{2}+36\left(2\alpha-1\right)e^{-2\alpha}+27e^{-4\alpha}\right]\\
 &  & C=\left(2\alpha-1+e^{-2\alpha}\right)\left(2\alpha-1\right)^{3}.
\end{eqnarray*}
Eq.\ (\ref{eq:bound_holmboe}) corroborates the  normal-mode result given in Appendix \ref{app:A}. 

The phase portrait of Holmboe instability, corresponding to an unstable combination of $\alpha$ and $J$, is shown in figure\ \ref{fig:Holmboe}(c). This phase portrait is slightly different 
from TC and KH instabilities, because $R_{NM} \neq 1$ in this case. Another feature of Holmboe instability is that, unlike  TC and KH instabilities, its phase-speed is non-zero at the equilibrium condition. This phase-speed is found to be
\begin{equation}
c_{1}=c_{2}=\frac{2J}{R_{NM}^{2}}=\frac{4J}{1-2\alpha+\sqrt{32\alpha J+\left(1-2\alpha\right)^{2}}}.
\label{eq:holmboe_ph_sp}
\end{equation}
In the limit of large $\alpha$ and $J$, the two phase-locked waves move with a speed of unity  to the positive $x$ direction.
\section{Conclusion}
Shear instability plays a crucial role in atmospheric and oceanic flows.  
%Although being studied for more than a century, 
%many aspects of shear instability has remained unclear - mainly because of the less intuitive approach taken to understand the problem. 
In the last 50 years, significant efforts have been made to develop a mechanistic understanding of shear instabilities.
 Using idealized velocity and density profiles, researchers have conjectured that the
resonant interaction between two counter-propagating linear
interfacial waves is the root cause behind exponentially growing instabilities in homogeneous and stratified shear layers. %\cite[]{holm1962,cair1979,caul1994,bain1994,carp2012}. 
Support for this claim has been provided by considering interacting vorticity and gravity waves of the normal-mode form.

In this paper we investigated the wave interaction problem in a generalized sense.  The governing equations (\ref{eq:27})-(\ref{eq:30}) of
hydrodynamic instability in idealized (broken-line profiles), homogeneous or density stratified, inviscid shear layers have been derived \emph{without} imposing the wave type, or the normal-mode waveform.  
We refer to this equation-set as  Wave Interaction Theory (WIT).
Using WIT we showed in  figures\ \ref{fig:Summaryfig}(a) and \ref{fig:Summaryfig}(b) that two counter-propagating linear
interfacial waves, having \emph{arbitrary} initial amplitudes and phases, eventually  \emph{resonate} (lock in phase and amplitude), provided they satisfy the N\&S condition (\ref{eq:necessary}). In \S \ref{subsub:eigenanalysis}
 we showed that the N\&S condition is basically  the criterion for normal-mode type instabilities. The waves which do not satisfy the N\&S condition  may exhibit non-normal instabilities, but will never resonate. In other words,  such waves will never lock in phase and undergo exponential growth. These facts have been demonstrated in figures  \ref{fig:manyfigs}(c) and \ref{fig:manyfigs}(d).  We considered three classic examples of shear instabilities - Rayleigh/Kelvin-Helmholtz (KH), Taylor-Caulfield (TC) and Holmboe, and derived their discrete spectrum non-modal stability equations by modifying the basic WIT equations (\ref{eq:27})-(\ref{eq:30}). These equations are respectively given by   (\ref{eq:khh1})-(\ref{eq:khh4}), (\ref{eq:T1})-(\ref{eq:T4}) and (\ref{eq:H1})-(\ref{eq:H4}).   For each  type of instability we validated the corresponding equation-set by showing that the N\&S condition  matches the predictions of the canonical normal-mode theory. 
%The discrete spectrum non-modal instabilities of 
%equation-sets (each set  corresponding to a classic instability) we formulated 
%Our contribution is in the formulation of the non-modal equation-sets for each type of instability mentioned in the previous paragraph. %These equation-sets describe the transient processes occurring in layered shear flows. 

 Although validation  is an important first step for checking the non-modal equation-sets,  the aim of this work is \emph{not} proving  well-known results using an alternative theory.  The real strength of  the non-modal formulation is in providing the opportunity to explore the transient dynamics. Non-orthogonal interaction between the two wave modes can lead to rapid transient growth. We show in figure \ref{fig:manyfigs}(b)  that depending on wavenumber and initial phase-shift,  non-modal gain can exceed corresponding modal-gain by several orders of magnitude. This implies that the flow may become  non-linear during the initial stages of flow development, leading to an early transition to turbulence.  Instability has been observed for wavenumbers which are deemed stable by the normal-mode theory. All these facts are shown for the example of KH instability; see  figures \ref{fig:manyfigs}(a)-(d). In order to be able to study the transient dynamics of Holmboe and TC instabilities, the analysis in \S \ref{sec:matstuff} needs to be modified in parts.
This is because the equations developed in \S \ref{sec:matstuff}  are restricted to systems whose non-modal equation-sets are \emph{homomorphic} to (\ref{eq:27})-(\ref{eq:30}).  Some examples of homomorphic equation-sets are:  (\ref{eq:khh1})-(\ref{eq:khh4}) representing KH instability,  (14a)-(14d) of \cite{heif1999} representing barotropic shear instability, and  (7a)-(7d) of \cite{davies1994} representing baroclinic instability of the Eady model. 

%The equations (\ref{eq:T1})-(\ref{eq:T4}) governing TC or (\ref{eq:H1})-(\ref{eq:H4}) governing Holmboe 

%The equations (\ref{eq:T1})-(\ref{eq:T4}) governing TC or (\ref{eq:H1})-(\ref{eq:H4}) governing Holmboe instability are not homomorphic to (\ref{eq:27})-(\ref{eq:30}). %Hence the analysis of \S \ref{sec:matstuff} needs to be modified to make it suitable for studying Holmboe and TC. 
%instability are not homomorphic to (\ref{eq:27})-(\ref{eq:30}). 

%. In \S \ref{subsec:non_mod} we developed the tools for studying transient dynamics  in a generalized sense, and  KH was chosen as an example; see  figures \ref{fig:manyfigs}(a)-(d). Transient growths of Holmboe and TC need to be studied in future.  
 %The  of KH, TC and Holmboe are respectively given by
% Resonance makes the interfacial waves, and therefore the shear layer, to  grow at an exponential rate.  

Another objective of our work is to provide an intuitive description of shear instabilities, and to find connection with other physical processes. How two waves can interact to produce instability has been discussed throughout the paper, and  schematically represented in figure \ref{fig:wave_int}. We observed an analogy between WIT equations and that governing the synchronization of two coupled harmonic oscillators.  On the basis of this analogy, we re-framed WIT as a non-linear dynamical system. The resonant configuration of the wave equations translated into a steady state configuration of the dynamical system. This dynamical system is of the source-sink type; the source and the sink nodes being the two equilibrium points. When interpreted in terms of  canonical linear stability theory, the source and the sink nodes respectively correspond to the decaying and the growing normal-modes of the discrete spectrum. The analogous two coupled harmonic oscillator system exhibits two normal modes of vibration - the in-phase and the anti-phase synchronization modes. The growing normal-mode of WIT is analogous to the in-phase synchronization mode, while the decaying normal-mode is analogous to the anti-phase synchronization mode.

%Probably the most important aspect of WIT is that it provides a non-modal description of instabilities occurring  in shear layers with idealized broken-line profiles. Non-modal instability  signifies 

% 
%  Although we have limited our study to the interaction between two waves,  the TC and Holmboe configurations actually  involve multiple wave interactions, which we have neglected.  Each density interface in these configurations supports two gravity waves, out of which only the counter-propagating wave has been considered. Since two wave interactions are sufficient to produce the normal mode characteristics of Holmboe and TC, the inclusion of the co-propagating gravity wave will only effect  the initial (non-modal)  stages of interaction. This effect may be non-trivial, and needs to be studied in future.  The WIT framework needs to be extended in future to include systems with multiple interfaces, where each interface can support multiple waves. 
 
 Although we studied WIT in the context of shear instabilities, the framework is actually derived from (\ref{eq:23})-(\ref{eq:26}), and is therefore applicable to both sheared and unsheared flows, as well as co-propagating and counter-propagating waves.  As an example, one can study the interaction between deep water surface gravity and internal gravity waves (which gives rise to barotropic and baroclinic modes of oscillations in lakes; see \cite{kundu2004} Pg.\ 259-261). The framework can also be extended to the weakly non-linear regime to study multiple-wave interactions leading to resonant triads \cite[]{wen1995,hill1996,jamali2003}. WIT can also be applied to understand liquid-jet instability \cite[]{jayz2000}.

Finally we focus on the implications of using broken-line profiles of velocity and/or density. These idealizations have allowed us to concentrate only on the discrete spectrum dynamics and understand hydrodynamic instability  in terms of interfacial wave interactions. 
 Real  profiles are always continuous, hence realistic analysis requires inclusion of the continuous spectrum. Using Green's function, \cite{heif2005} have shown that the continuous spectrum dynamics in a smooth homogeneous shear layer  can be understood in terms of infinite number of interacting  vorticity (Rossby edge) waves. \cite{harnik2008} used the same approach to understand the normal-mode continuous spectrum of smooth stratified shear layers. These studies indicate that the growth process is  better understood through the ``Orr mechanism'' of shearing of waves \cite[]{orr1907,heif2005}, than edge wave interactions. 
This implies that the intuition brought by WIT formulation may become less apparent in continuous systems. However the applicability of WIT in continuous systems can only be  properly judged when  the current version is modified to include the effects of a continuous spectrum. We speculate that the analogy between WIT and the synchronization of coupled harmonic oscillators can be exploited to obtain an intuitive understanding of  continuous spectrum dynamics. Our speculation is based on the work of \cite{balm2000}, who studied the Kuramoto model in the continuous limit, and observed superficial similarities between oscillator synchronization and instabilities in ideal plasmas and inviscid fluids. 
 
%  Summarizing,  our two-wave-interaction problem gets modified into an infinite-wave-interaction problem in the case of continuous profiles, however the underlying mechanism remains the same. 
 
 \begin{acknowledgments}
We would like to acknowledge the useful comments and suggestions of Prof. Neil Balmforth of UBC, Prof. Kraig Winters of UCSD, Prof. Michael McIntrye, Prof. Colm Caulfield and Prof. Peter Haynes of the University of Cambridge (DAMTP), Prof. Eyal Heifetz of Tel Aviv University, Prof. Ishan Sharma and Prof. Mahendra Verma of IIT Kanpur, and the anonymous referees.   
\end{acknowledgments}
 
 \appendix

 \section{Stokes' theorem applied to a vorticity interface}
\label{app:B}

\begin{figure}
\centering
\includegraphics[scale=0.13]{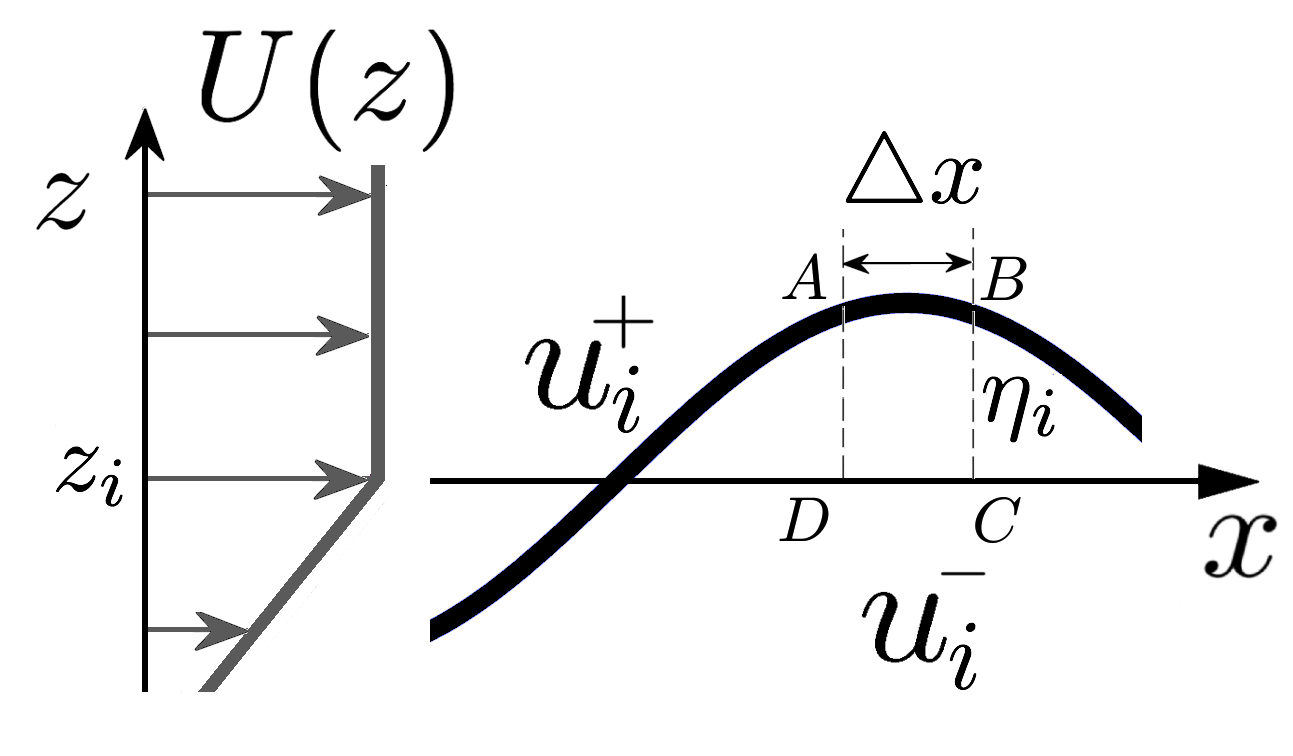}
 \caption{Schematic of a vorticity interface - left half shows unperturbed velocity field, while the right half depicts infinitesimal interfacial displacement.}
\label{fig:Appen}
\end{figure}

Stokes' theorem relates the surface integral of the curl of a vector (velocity in this case) field $\vec{u}$ over a surface $A$  to the line integral of the vector field over its boundary $\delta A$:
\begin{equation}
\oint_{\delta A}\vec{u} \cdot d\vec{l}=\iint_{A}\left(\nabla\times\vec{u}\right) \cdot d\vec{A}.
\label{eq:app_stokes}
\end{equation}
\cite{holm1962} used this theorem to relate the interfacial displacement $\eta_{i}$ with the difference in velocity perturbation ($u_{i}^{+}-u_{i}^{-}$) produced at a vorticity interface;  see   (\ref{eq:Stokes}). This equation is referred to as ``Eq.\ (3.2)'' in his paper. However, the relevant steps required to derive this equation were not  provided. In order to understand how  (\ref{eq:Stokes}) is obtained, we first graphically describe the problem in  figure \ref{fig:Appen}.  The background velocity is such that the flow is irrotational when $z>z_{i}$, and has a constant vorticity, say $S$, when $z \leq z_{i}$. When the interface is disturbed by an infinitesimal displacement  $\eta_{i}$ (solid black curve in figure\ \ref{fig:Appen}), the velocity field also changes slightly - the perturbation velocity in the upper layer ($z>z_{i}$) becomes  $u_{i}^{+}$ and that in the lower layer ($z \leq z_{i}$) becomes $u_{i}^{-}$. 

Let us consider a circuit A-B-C-D. Applying Stokes' theorem, we obtain
\begin{equation}
(u_{i}^{+}-u_{i}^{-})\triangle x=S \cdot A,
\end{equation}
where $A= \eta_{i} \cdot \triangle x$ is the area of A-B-C-D, and $S=\nabla\times\vec{u}$ is the vorticity in this area. Therefore we obtain 
\begin{equation}
u_{i}^{+}-u_{i}^{-}=S \cdot \eta_{i},
\end{equation}
which is basically (\ref{eq:Stokes}).

\section{Normal mode form of Holmboe instability}
\label{app:A}

Both interfaces in the Holmboe profile (\ref{eq:holm1}) individually satisfy the kinematic condition:
 \begin{eqnarray}
 &  & \frac{\partial\eta_{1}}{\partial t}=\frac{\partial}{\partial x}\left(e^{-\alpha}\psi_{2}+\frac{1-2\alpha}{2\alpha}\eta_{1}\right)\\
  &  & \frac{\partial\eta_{2}}{\partial t}=\frac{\partial}{\partial x}\left(\psi_{2}+\frac{e^{-\alpha}}{2\alpha}\eta_{1}\right),
\end{eqnarray}
where $\psi_{2}$ is the stream function perturbation at the lower interface. This interface being a density interface also satisfies the
dynamic condition:
\begin{equation}
 \frac{\partial\psi_{2}}{\partial x}=\frac{J}{\alpha}\frac{\partial\eta_{2}}{\partial x}.
 \end{equation}

We assume the perturbations to be of normal-mode form:
$\eta_{1}=\Re\{\hat{\eta}_{1} e^{\ii\alpha(x-ct)}\}$, $\eta_{2}=\Re\{\hat{\eta}_{2} e^{\ii\alpha(x-ct)}\}$, and $\psi_{2}=\Re\{\hat{\psi}_{2} e^{\ii\alpha(x-ct)}\}$. Here the wave speed $c$ is generally  complex.
Defining $\hat{\varsigma}=\left[\begin{array}{ccc}
\hat{\psi}_{2} & \hat{\eta}_{2} & \hat{\eta}_{1}\end{array}\right]^{T}$, we obtain the following eigenvalue problem:
\begin{equation}
\left(M+cI\right)\hat{\varsigma}=0,
\label{eq:app2}
\end{equation}
where
\begin{equation}
M=\left[\begin{array}{ccc}
0 & J/\alpha & 0\\
1 & 0 & e^{-\alpha}/(2\alpha)\\
e^{-\alpha} & 0 & (1-2\alpha)/(2\alpha)
\end{array}\right].
\end{equation}
 Eq.\ (\ref{eq:app2}) 
 generates the following characteristic polynomial:
\begin{equation}
c^{3}+\left(\frac{1-2\alpha}{2\alpha}\right)c^{2}-\frac{J}{\alpha}c-J\left(\frac{1-2\alpha}{2\alpha^{2}}\right)+J\frac{e^{-2\alpha}}{2\alpha^{2}}=0.
\end{equation}
This equation produces complex conjugate roots only when the discriminant is negative.  Since the presence of complex roots signify normal-mode instability, negative values of the discriminant is of our interest. The discriminant (D) in this case is given by:
\begin{equation}
D=16\alpha^{2}J^{2}-\alpha J\left[8\left(2\alpha-1\right)^{2}+36e^{-2\alpha}\left(2\alpha-1\right)+27e^{-4\alpha}\right]-\left(1-2\alpha\right)^{3}\left(2\alpha-1+e^{-2\alpha}\right).
\end{equation}
Imposing the condition $D<0$, we find
\begin{equation}
\frac{1}{2A}\left(-B-\sqrt{B^{2}-4AC}\right)\leq J\leq\frac{1}{2A}\left(-B+\sqrt{B^{2}-4AC}\right),
\label{eq:bbbb}
\end{equation}
where
\begin{eqnarray*}
 &  & A=16\alpha^{2}\\
 &  & B=-\alpha\left[8\left(2\alpha-1\right)^{2}+36\left(2\alpha-1\right)e^{-2\alpha}+27e^{-4\alpha}\right]\\
 &  & C=\left(2\alpha-1+e^{-2\alpha}\right)\left(2\alpha-1\right)^{3}.
\end{eqnarray*}
Thus Holmboe instability occurs only when the condition  in  (\ref{eq:bbbb})  is satisfied.

\bibliographystyle{jfm}

%\bibliography{paper1}

\begin{thebibliography}{44}
\expandafter\ifx\csname natexlab\endcsname\relax\def\natexlab#1{#1}\fi

\bibitem[Baines \& Mitsudera(1994)]{bain1994}
{\sc Baines, P. G. \& Mitsudera, H.} 1994 On the mechanism of shear flow
  instabilities. {\em J. Fluid Mech.\/} {\bf 276}, 327--342.

\bibitem[Bakas \& Ioannou(2011)]{bakas2009}
{\sc Bakas, N. A. \& Ioannou, P. J.} 2009 Modal and nonmodal growths of inviscid planar
 perturbations in shear flows with a free surface. {\em Phys. Fluids\/} {\bf 21}~(2),
  024102.
  

\bibitem[Balmforth \& Sassi(2000)]{balm2000}
{\sc Balmforth, N. J. \& Sassi, R.} 2000 A shocking display of synchrony.
 {\em Phys. D\/} {\bf 143}, 21--55.


\bibitem[Bretherton(1966)]{bret1966}
{\sc Bretherton, F.~P.} 1966 Baroclinic instability and the short wavelength
  cut-off in terms of potential vorticity. {\em Q. J. Roy. Meteor. Soc.\/} {\bf
  92}~(393), 335--345.

\bibitem[Cairns(1979)]{cair1979}
{\sc Cairns, R. A.} 1979 The role of negative energy waves in some instabilities
  of parallel flows. {\em J. Fluid Mech.\/} {\bf 92}, 1--14.

\bibitem[Carpenter {\em et~al.\/}(2013)Carpenter, Tedford, Heifetz \&
  Lawrence]{carp2012}
{\sc Carpenter, J. R., Tedford, E. W., Heifetz, E. \& Lawrence,
  G. A.} 2013 Instability in stratified shear flow: Review of a physical
  interpretation based on interacting waves. {\em Appl. Mech. Rev.\/}
  {\bf 64}~(6), 060801-17.

\bibitem[Caulfield(1994)]{caul1994}
{\sc Caulfield, C. P.} 1994 Multiple linear instability of layered stratified
  shear flow. {\em J. Fluid Mech.\/} {\bf 258}, 255--285.

\bibitem[Caulfield {\em et~al.\/}(1995)Caulfield, Peltier, Yoshida \&
  Ohtani]{caul1995}
{\sc Caulfield, C. P., Peltier, W. R., Yoshida, S. \& Ohtani, M.} 1995 An
  experimental investigation of the instability of a shear flow with
  multilayered density stratification. {\em Phys. Fluids\/} {\bf 7},
  3028--3041.

\bibitem[Constantinou \& Ioannou(2011)]{const2011}
{\sc Constantinou, N. C. \& Ioannou, P. J.} 2011 Optimal excitation of
  two dimensional Holmboe instabilities. {\em Phys. Fluids\/} {\bf 23}~(7),
  074102.

\bibitem[Davies \& Bishop(1994)]{davies1994}
{\sc Davies, H. C. \& Bishop, C. H.} 1994 Eady edge waves and rapid development. {\em
  J. Atmos. Sci.\/} {\bf 51}~(13), 1930--1946.


\bibitem[Drazin \& Howard(1966)]{draz1966}
{\sc Drazin, P. G. \& Howard, L. N.} 1966 {\em Hydrodynamic stability of parallel flow of inviscid fluid.\/}
{\bf 9}, Academic Press.

\bibitem[Drazin \& Reid(2004)]{draz1982}
{\sc Drazin, P. G. \& Reid, W. H.} 2004 {\em {H}ydrodynamic {S}tability\/}, 2nd
  edn. Cambridge University Press.

\bibitem[Farrell(1984)]{farrell1984modal}
{\sc Farrell, B.} 1984 Modal and non-modal baroclinic waves. {\em J. Atmos.
  Sci.\/} {\bf 41}~(4), 668--673.

\bibitem[Farrell \& Ioannou(1996)]{farrell1996}
{\sc Farrell, B. F. \& Ioannou, P. J.} 1996 Generalized stability theory. Part I:
  Autonomous operators. {\em J. Atmos. Sci.\/} {\bf 53}~(14), 2025--2040.

\bibitem[Goldstein(1931)]{gold1931}
{\sc Goldstein, S.} 1931 On the stability of superposed streams of fluids of
  different densities. {\em Proc. R. Soc. Lond. A\/} {\bf 132}, 524--548.

\bibitem[Guha, Rahmani \& Lawrence(2013)]{guha2012}
{\sc Guha, A., Rahmani, M. \& Lawrence, G. A.} 2013 Evolution of a
  barotropic shear layer into elliptical vortices. {\em Phys. Rev. E\/} {\bf
  87}, 013020.

\bibitem[Harnik {\em et~al.\/}(2008)Harnik, Heifetz, Umurhan \&
  Lott]{harnik2008}
{\sc Harnik, N., Heifetz, E., Umurhan, O. M. \& Lott, F.} 2008 A
  buoyancy-vorticity wave interaction approach to stratified shear flow. {\em
  J. Atmos. Sci.\/} {\bf 65}~(8), 2615--2630.

\bibitem[Heifetz, Bishop \& Alpert(1999)]{heif1999}
{\sc Heifetz, E., Bishop, C. H. \& Alpert, P.} 1999 Counter-propagating {R}ossby
  waves in the barotropic {R}ayleigh model of shear instability. {\em Q. J. R.
  Meteorol. Soc.\/} {\bf 125}~(560), 2835--2853.

\bibitem[Heifetz {\em et~al.\/}(2004)Heifetz, Bishop, Hoskins \&
  Methven]{heifetz2004counter}
{\sc Heifetz, E., Bishop, C. H., Hoskins, B. J. \& Methven, J.} 2004 The
  counter-propagating rossby-wave perspective on baroclinic instability. I:
  Mathematical basis. {\em Q. J. Roy. Meteor.
  Soc.\/} {\bf 130}~(596), 211--231.

\bibitem[Heifetz \& Methven(2005)]{heif2005}
{\sc Heifetz, E. \& Methven, J.} 2005 Relating optimal growth to
  counterpropagating {R}ossby waves in shear instability. {\em Phys. Fluids\/}
  {\bf 17}~(6), 064107.


\bibitem[Hill \& Foda(1996)]{hill1996}
{\sc Hill, D. F. \& Foda, M. A.} 1996 Subharmonic resonance of short internal standing waves by progressive surface waves.
 {\em J. Fluid Mech.\/} {\bf 321}, 217--234.

\bibitem[Holmboe(1962)]{holm1962}
{\sc Holmboe, J.} 1962 On the behavior of symmetric waves in stratified shear
  layers. {\em Geofys. Publ.\/} {\bf 24}, 67--112.

\bibitem[Hoskins, McIntyre \& Robertson(1985)]{hosk1985}
{\sc Hoskins, B. J., McIntyre, M. E. \& Robertson, A. W.} 1985 On the use and
  significance of isentropic potential vorticity maps. {\em Q. J. Roy. Meteor.
  Soc.\/} {\bf 111}~(470), 877--946.


\bibitem[Howard \& Maslowe(1973)]{hm73}
{\sc Howard, L. N. \& Maslowe, S. A.} 1973 Stability of stratified shear flows.
 {\em Bound.-Layer Meteor.\/}
  {\bf 4}~(1--4), 511--523.

\bibitem[Jamali, Seymour \& Lawrence(2003)]{jamali2003}
{\sc Jamali, M., Seymour, B. \& Lawrence, G. A.} 2003 Asymptotic analysis of a surface-interfacial wave interaction. 
{\em Phys. Fluids\/} {\bf
  15}~(1), 47--55.

\bibitem[Jazayeri \& Li(2000)]{jayz2000}
{\sc Jazayeri,S. A. \& Li, X.} 2000 Nonlinear instability of plane liquid sheets. {\em J. Fluid Mech.\/} {\bf
  406}, 281--308.

\bibitem[Kundu \& Cohen(2004)]{kundu2004}
{\sc Kundu, P. K. \& Cohen, I. M.} 2004 {\em Fluid Mechanics\/}. Elsevier.


\bibitem[Lawrence, Browand \& Redekopp(1991)]{law1991}
 {\sc Lawrence, G. A.,  Browand, F. K. \& Redekopp, L. G.} 1991 The stability of a sheared density interface. 
{\em Phys. Fluids\/} {\bf3}~(10), 2360--2370.

\bibitem[Lee \& Caulfield(2001)]{lc2001}
{\sc Lee, V. \& Caulfield, C. P.} 2001 Nonlinear evolution of a layered stratified shear flow.
 {\em Dyn. Atmos. Oceans\/}
  {\bf 34}~(2), 103--124.

\bibitem[Lindzen(1988)]{lind1988}
{\sc Lindzen, R. S.} 1988 Instability of plane parallel shear flow (toward a
  mechanistic picture of how it works). {\em Pure Appl. Geophys.\/} {\bf
  126}~(1), 103--121.
  
  \bibitem[Miles(1957)]{miles1957}
{\sc Miles, J. W.} 1957 On the generation of surface waves 
by shear flows. {\em J. Fluid Mech.\/} {\bf
  3}, 185--204.
  

\bibitem[Orr(1907)]{orr1907}
{\sc Orr, W. M.~F.} 1907 Stability or instability of the steady motions of a
  perfect liquid and of a viscous liquid. {\em Proc. Roy. Irish Acad.\/} {\bf
  A}~(27), 9--138.

%\bibitem[Pouliquen, Chomaz \& Huerre(1994)]{poq1994}
%{\sc Pouliquen, O., Chomaz, J., \& Huerre, P.} 1994 Propagating Holmboe waves at the interface
%between two immiscible fluids. {\em J. Fluid Mech.\/} {\bf 266} 277–-302.


\bibitem[Pikovsky, Rosenblum \& Kurths(2001)]{PIK01}
{\sc Pikovsky, A., Rosenblum, M. G. \& Kurths, J.} 2001 {\em Synchronization, A Universal
 Concept in Nonlinear Sciences\/}, Cambridge University Press.

\bibitem[Rayleigh(1880)]{rayl1880}
{\sc Rayleigh, J. W. S.} 1880 On the stability, or instability, of certain fluid
  motions. {\em Proc. Lond. Math. Soc.\/} {\bf 12}, 57--70.
  
\bibitem[Rabinovich et al.(2011)]{rabi2011}
{\sc Rabinovich, A., Umurhan, O. M., Harnik, N., Lott, F. \& Heifetz, E.} 2011 Vorticity inversion and action-at-a-distance instability in stably stratified shear flow.
 {\em J. Fluid Mech.\/} {\bf
  670}, 301--325.

\bibitem[Schmid \& Henningson(2001)]{schmid2001stability}
{\sc Schmid, P. J. \& Henningson, D. S.} 2001 {\em Stability and transition in
  shear flows\/},  {\bf 142}. Springer Verlag.

\bibitem[Sutherland(2010)]{suth2010}
{\sc Sutherland, B. R.} 2010 {\em Internal gravity waves\/}, Cambridge
  University Press.

\bibitem[Taylor(1931)]{tayl1931}
{\sc Taylor, G. I.} 1931 Effect of variation in density on the stability of
  superposed streams of fluid. {\em Proc. R. Soc. Lond. A\/} {\bf 132},
  499--523.

\bibitem[Tedford, Pieters \& Lawrence(2009)]{ted2009}
{\sc Tedford, E. W., Pieters, R. \& Lawrence, G. A.} 2009 Symmetric Holmboe
  instabilities in a laboratory exchange flow. {\em J. Fluid Mech.\/} {\bf
  636}, 137--153.
 
 \bibitem[Tedford et al.(2009)]{ted2009a}
{\sc Tedford, E. W., Carpenter, J.R., Pawlowicz, R., Pieters, R. \& Lawrence, G. A.} 2009 
  Observation and analysis of shear instability in the Fraser River estuary. {\em J. Geophys. Res.\/} {\bf
  114}, C11.
  
\bibitem[Thorpe(1973)]{thor1973}
{\sc Thorpe, S. A.} 1973 Experiments on instability and turbulence in a
  stratified shear flow. {\em J. Fluid Mech.\/} {\bf 61}, 731--751.

\bibitem[Trefethen {\em et~al.\/}(1993)Trefethen, Trefethen, Reddy \&
  Driscoll]{tref1993}
{\sc Trefethen, L. N., Trefethen, A. E., Reddy, S. C. \& Driscoll, T. A.} 1993
  Hydrodynamic stability without eigenvalues. {\em Science\/} {\bf 261},
  578--584.

\bibitem[Turner(1973)]{turn1973}
{\sc Turner, J. S.} 1973 {\em Buoyancy Effects in Fluids\/}, Cambridge University Press.

\bibitem[Wen(1995)]{wen1995}
{\sc Wen, F.} 1995 Resonant generation of internal waves on the soft sea bed by a surface water wave. 
{\em Phys. Fluids\/} {\bf
  7}~(8), 1915--1922.


\end{thebibliography}

\end{document}